\documentclass[3p]{elsarticle}

\usepackage{bbm}
\usepackage{amssymb}
\usepackage{ifpdf}
\usepackage{color}
\usepackage{graphicx,epsfig}

\definecolor{myblue}{rgb}{0.2,0.2,0.8}
\definecolor{myred}{rgb}{1,0.,0.3}

\usepackage[colorlinks=true,citecolor=myblue,linkcolor=myred]{hyperref}

\usepackage{MnSymbol}
\usepackage{graphicx,amsmath,amssymb,dsfont}
\usepackage{color}
\usepackage{psfrag}
\usepackage{bm}
\newcommand{\ket}[1]{| #1 \rangle}

\newcommand{\beq}{\begin{eqnarray}}
\newcommand{\eeq}{\end{eqnarray}}

\newcommand{\vertvbket}{| \raisebox{-1.5 pt}{\includegraphics[scale=0.042]{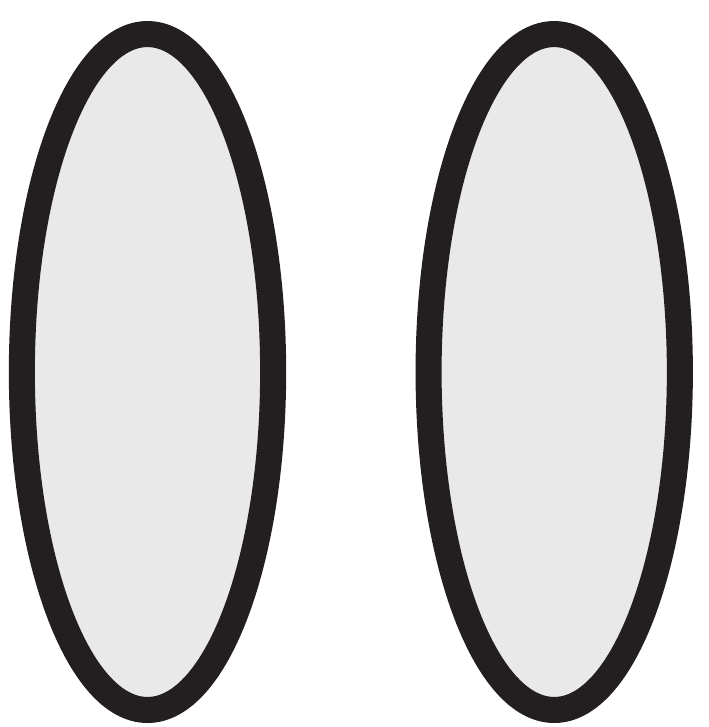}} \rangle}
\newcommand{\horvbbra}{\langle \raisebox{-1.5 pt}{\includegraphics[scale=0.04]{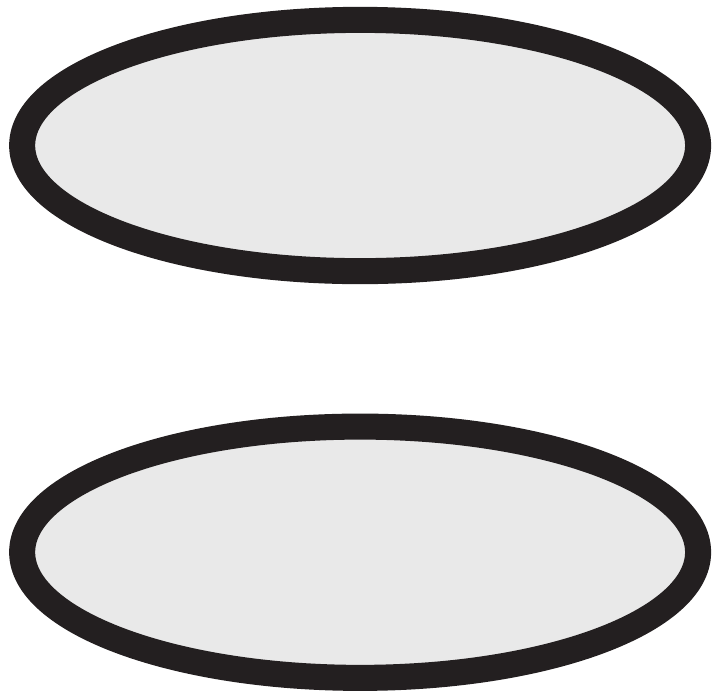}} |}
\newcommand{\vertvbbra}{\langle \raisebox{-1.5 pt}{\includegraphics[scale=0.04]{vdimer.pdf}} |}
\newcommand{\horvbket}{| \raisebox{-1.5 pt}{\includegraphics[scale=0.04]{hdimer.pdf}} \rangle}

\begin{document}
\title{Two-dimensional Lattice Gauge Theories\\ with Superconducting Quantum Circuits}

\author[iqoqi]{D. Marcos}
\ead{david.marcos@me.com}
\author[bern]{P. Widmer}
\author[strasbourg]{E. Rico}
\author[jqi,cpark]{M. Hafezi}
\author[iqoqivienna]{P. Rabl}
\author[bern]{U.-J. Wiese}
\author[iqoqi,innsbruckuni]{P. Zoller}

\address[iqoqi]{Institute for Quantum Optics and Quantum Information of the Austrian Academy of Sciences, A-6020 Innsbruck, Austria}
\address[bern]{Albert Einstein Center, Institute for Theoretical Physics, Bern University, CH-3012, Bern, Switzerland}
\address[strasbourg]{IPCMS (UMR 7504) and ISIS (UMR 7006), University of Strasbourg and CNRS, 67000 Strasbourg, France}
\address[jqi]{Joint Quantum Institute, NIST/University of Maryland, College Park 20742, USA}
\address[cpark]{ECE Institute for Research in Electronics and Applied Physics, University of Maryland, College Park, MD 20742, USA}
\address[iqoqivienna]{Institute of Atomic and Subatomic Physics, TU Wien, Stadionallee 2, 1020 Wien, Austria}
\address[innsbruckuni]{Institute for Theoretical Physics, University of Innsbruck, A-6020 Innsbruck, Austria}

\begin{abstract}
A quantum simulator of $U(1)$ lattice gauge theories can be implemented with superconducting circuits. This allows the investigation of confined and deconfined phases in quantum link models, and of valence bond solid and spin liquid phases in quantum dimer models. Fractionalized confining strings and the real-time dynamics of quantum phase transitions are accessible as well. Here we show how state-of-the-art superconducting technology allows us to simulate these phenomena in relatively small circuit lattices. By exploiting the strong non-linear couplings between quantized excitations emerging when superconducting qubits are coupled, we show how to engineer gauge invariant Hamiltonians, including ring-exchange and four-body Ising interactions. We demonstrate that, despite decoherence and disorder effects, minimal circuit instances allow us to investigate properties such as the dynamics of electric flux strings, signaling confinement in gauge invariant field theories. The experimental realization of these models in larger superconducting circuits could address open questions beyond current computational capability.
\end{abstract}

\maketitle

\section{Introduction}

Since the pioneering experiments showing quantized coherent excitations in electrical circuits \cite{Nakamura,vanderWal}, superconducting circuits including Josephson junctions are playing a fundamental role to demonstrate quantum effects at a mesoscopic level and, remarkably, in quantum information processing. The enormous recent progress in this field comprises, for example, the realization of quantum teleportation \cite{Wallraff13} and complex two- and three-qubit algorithms, including number factoring and quantum error correction \cite{DiCarlo09, Martinis12, Martinis11, Wallraff12, Schoelkopf12}. From the viewpoint of analog quantum simulation, the large coherence times and non-linearities achieved with superconducting qubits \cite{Schoelkopf11, Steffen12, Kirchmair13, DevoretSchoelkopfReview} have opened frontiers towards the simulation of Hubbard models with photonic excitations and, as a by-product, the emulation of classical static fields in circuit lattices \cite{HouckTureciKoch, Koch10, Houck12}. 

A new perspective in quantum simulation is to mimic fundamental interactions, such as those arising in field theories \cite{Preskill}, and in particular, lattice gauge theories \cite{WieseReview}. In elementary particle physics, dynamical quantum gauge fields mediate fundamental interactions \cite{Wilson, Kogut-Susskind, Gattringer}. In condensed matter systems such as spin liquids, dimer models, and presumably in high-temperature superconductors, gauge fields emerge as relevant low-energy degrees of freedom \cite{KogutSpinsRMP, Wen, LeeNagaosaWen, Lacroix, Balents}. Solving these theories is, however, fundamentally challenging. Classical simulations typically rely on Monte Carlo methods which may suffer from severe sign problems, which imply that real-time dynamics and certain exotic phases are so far out of reach. The quantum simulation of {\it dynamical} gauge fields is  thus attracting a great deal of interest, giving rise to a variety of recent proposals, mainly based on cold atoms in optical lattices \cite{Buchler05, Weimer10, Kapit11, Zohar11, Banerjee12, Zohar12, Zohar13, Lewenstein12, Banerjee13, Zohar13b, Lewenstein13, Zohar13c}.

Here we show how different gauge invariant models can be simulated with superconducting circuits. This platform offers on-chip highly-tunable couplings, and local control over basic modules that can be interconnected, enabling --- in principle --- scalability. Specifically, in this work we focus our attention on two-dimensional $U(1)$ gauge theories, and show how ring-exchange interactions, present in dimer models, and plaquette terms arising in lattice gauge theories, can be engineered with quantum circuits under realistic dissipative conditions. We will illustrate this by constructing gauge invariant models in a superconducting-circuit square lattice. As we will show, even in the presence of excitation loss and disorder, distinctive features of the gauge theory, such as confinement and string dynamics, can be observed in relatively small circuit lattices. The implementation of these gauge invariant interactions generalizes previous proposals based on cold atoms \cite{Buchler05, Weimer10, Kapit11, Zohar11, Banerjee12, Zohar12, Zohar13, Lewenstein12, Banerjee13, Zohar13b, Lewenstein13, Zohar13c, Glaetzle}, as well as pioneering studies in this area with Josephson-junction arrays \cite{Ioffe}, trapped ions \cite{Hauke13}, and superconducting circuits \cite{Marcos13}.

To quantum simulate dynamical gauge fields, we use the framework of quantum link models \cite{Horn81, Orland90, Wiese97}. In this formulation, the gauge field is represented by quantum degrees of freedom residing on the links that connect neighboring lattice sites. In contrast to Wilson's lattice gauge theory \cite{Wilson, Kogut-Susskind}, quantum link models have a finite-dimensional Hilbert space per link, and provide an alternative non-perturbative regularization of gauge theories. This, on the one hand, leads to new theories beyond the Wilson framework, and, on the other hand, allows us to address the standard gauge field theories relevant in particle physics. For example, quantum chromodynamics (QCD) emerges from an $SU(3)$ invariant quantum link model by dimensional reduction \cite{Bro99}. In this framework, continuously varying gluon fields are not put in by hand, but emerge dynamically as collective excitations of discrete quantum link degrees of freedom, and chiral quarks can be incorporated naturally as domain wall fermions. Quantum electrodynamics and other gauge field theories relevant in particle
physics can be regularized with quantum links along the same lines. Here we focus our attention on the simplest $U(1)$ lattice gauge theories that can be realized with quantum links. While they are not directly connected with particle physics, they share qualitative features with QCD, including the
existence of confining flux strings. In addition, they are of interest in the context of the condensed matter physics in strongly correlated electron
systems.

For a $U(1)$ quantum link model, the link degrees of freedom may be represented by spin $S=\frac{1}{2}$ operators. Quantum dimer models have the same Hamiltonian as the $U(1)$ quantum link model, but operate in a static background of ``electric'' charges. Upon doping, quantum dimer models may realize Anderson's resonating valence bond scenario of high-temperature superconductivity \cite{AndersonRVB}. In this case, confinement manifests itself in valence bond solid phases, while deconfinement is associated with quantum spin liquids.
Confinement is characterized by the energy of the electric flux strings that connect charge and anticharge, and whose energy is proportional to the string length. In quantum link and quantum dimer models the strings fractionalize into strands of electric flux $\frac{1}{2}$ \cite{BanerjeeU1, BanerjeeProceedings} and $\frac{1}{4}$ \cite{BanerjeeArxiv14}, respectively. Of specific interest in the context of quantum simulation are dynamical properties, such as the evolution after a quench \cite{Trotzky}. In our lattice gauge theory, the time evolution of the confining strings is beyond current computational capability for relatively small lattices, and as we show below, could be addressed with a quantum simulator based on superconducting circuits. In particular, it would be interesting to investigate how an initially prepared confining string separates into fractionalized strands as a function of time, a process that is also relevant from a condensed matter perspective in the context of quantum dimer models. Although here we concentrate on small superconducting-circuit lattices that can be built with current superconducting-circuit technology, in the future, larger systems could be built to investigate subtle aspects of the string dynamics, both at the roughening transition and near a bulk phase transition, which can be captured by a low-energy effective string theory \cite{BanerjeeU1, BanerjeeProceedings}. In this sense, the proposed devices can be used to study ``{\it string theory on a chip}''.

The paper is organized as follows. In section \ref{QLMsec} we introduce quantum link and quantum dimer models, emphasizing their gauge symmetry. We construct the corresponding Hamiltonians and discuss associated phenomena, in particular, the dynamics of confining strings. In section \ref{SCimplementationSec} we show how the gauge invariant models of interest can be simulated with a superconducting-circuit architecture. Specifically, we analyze in detail the building blocks that compose the circuit lattice, and demonstrate how, for realistic parameters, the system can be tuned via external magnetic fields to give access to different parameter regimes, and thus the corresponding phases of the model. In section \ref{1plaquetteSec} we propose a minimal experiment to demonstrate ring-exchange dynamics in a single plaquette. In section \ref{2plaquettesSec} we study the physics associated with the competing energy scales of our model. In particular, we show how a bulk phase transition manifests itself in the behavior of a particular lattice state, and discuss the physics associated with electric flux strings. Our simulations of minimal instances pave the way towards experiments on small lattices to demonstrate dynamical effects in equilibrium and out-of-equilibrium gauge systems, which have been out of reach so far. In section \ref{ConclusionsSec} we present our conclusions and discuss possible directions for future developments.

\begin{figure*}[t]
\centering
\includegraphics[width=\textwidth]{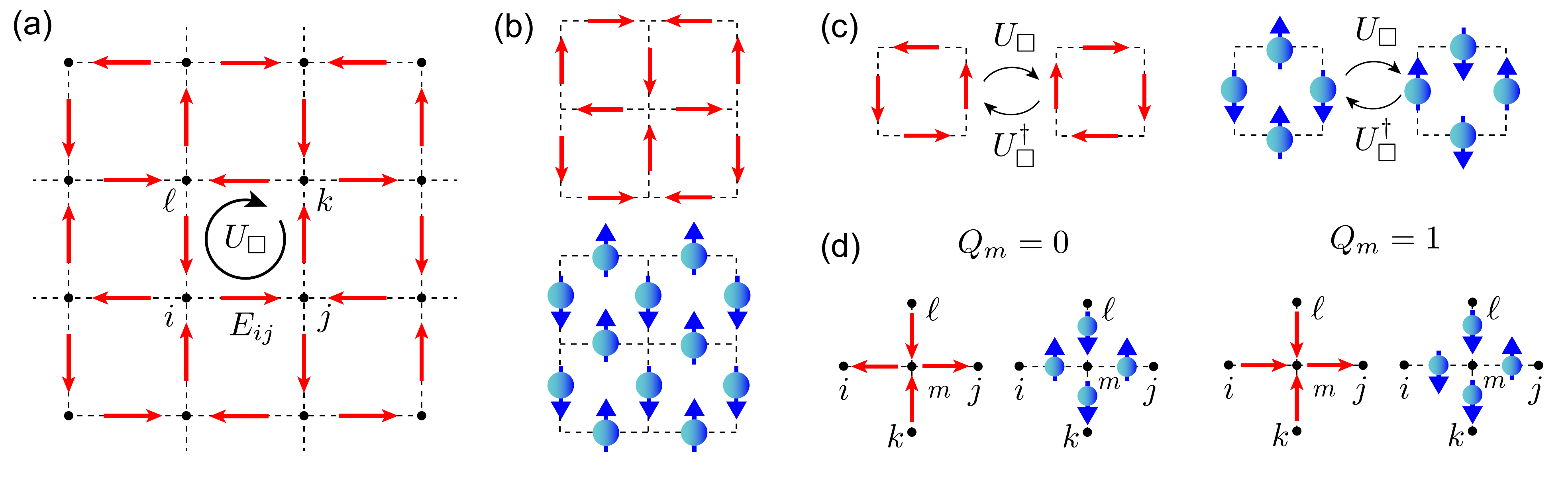}
\caption{ (Color online). (a) In a $U(1)$ lattice gauge theory, the electric field is represented by operators $E_{ij}$ that live on the links of a (two-dimensional) lattice.  An eigenstate $|e_{ij}\rangle$ of the electric field operator $E_{ij}$ is represented by a flux arrow from site $i$ to the neighboring site $j$. The plaquette operators $U_\square=U_{ij}U_{jk}U_{k\ell}U_{\ell i}$ act on the four electric flux states around a plaquette. (b) Mapping between an electric flux configuration and the corresponding spin states of the $S=\frac{1}{2}$ quantum link model.  (c) Action of the plaquette operators on the electric flux and spin $S=\frac{1}{2}$ representation. (d) Illustration of the Gauss law.}
\label{fig:general}
\end{figure*}

\section{Quantum link and quantum dimer models} \label{QLMsec}

In this work we consider the implementation of various $U(1)$ gauge theories on a two-dimensional lattice, using the quantum link model (QLM) formulation of lattice gauge theories. As already outlined in the Introduction, QLMs are lattice gauge theories with a finite-dimensional Hilbert space per link, which makes them ideally suited for quantum simulation. Moreover, prominent models in condensed matter physics, such as quantum spin ice or quantum dimer models, naturally fall in this theoretical framework \cite{Lacroix}. The purpose of this section is to establish the main concepts and a common notation used in the later parts of the paper. For an introduction to Wilson's standard formulation of lattice gauge theories the reader is referred to \cite{Wilson, Kogut-Susskind, Gattringer, KogutSpinsRMP}.

\subsection{$U(1)$ Quantum link models}

In the Hamiltonian formulation of Wilson-type Abelian lattice gauge theories --- such as quantum electrodynamics (QED) --- the dynamical gauge fields are represented by variables $U_{ij} = \exp({\rm i} \varphi_{ij}) \in U(1)$ that live on the links between two neighboring lattice sites $i$ and $j$.  Here $\varphi_{ij} = \int_i^j d\vec l \cdot \vec A$ corresponds to the phase accumulated by a charged particle moving from $i$ to $j$ in the presence of a vector potential $\vec A$. Associated to each link variable, there is a canonically conjugate electric flux operator $E_{ij} = - i \partial_{\varphi_{ij}}$ [see Fig.~\ref{fig:general}(a)], which obeys the commutation relations 
\begin{equation}
\label{commutatorEU}
[E_{ij},U_{ij}] = U_{ij}, \quad [E_{ij},U^\dagger_{ij}] = - U^\dagger_{ij}.
\end{equation}
In Wilson's lattice gauge theory the operator $E_{ij}$ acts on an infinite-dimensional local link Hilbert space, with eigenstates $E_{ij} |e_{ij}\rangle = e_{ij}|e_{ij}\rangle$ and eigenvalues $e_{ij} \in \mathbb{Z}$. The commutation relations (\ref{commutatorEU}) imply that $U_{ij}$ and $U^\dagger_{ij}$ act as raising and lowering operators of the electric flux $e_{ij}$, respectively. Here we use the convention that positive eigenvalues, $e_{ij}>0$, correspond to a flux from site $i$ to site $j$.  In the lattice formulation, the dynamics of the gauge fields is described by a Hamiltonian of the form
\begin{equation} \label{eq:HWilson}
\begin{split}
H &= \frac{g^2}{2}\sum_{\langle ij\rangle} E_{ij}^2 -\frac{1}{4g^2} \sum_{\square} \left( U_\square + U_{\square}^\dagger \right), \\
U_\square &= U_{ij}U_{jk}U_{k\ell}U_{\ell i} = \exp\{{\rm i} (\varphi_{ij} + \varphi_{jk} + \varphi_{k\ell} + \varphi_{\ell i})\} = \exp\{{\rm i} \Phi \}.
\end{split}
\end{equation}  
Here $\langle i,j\rangle$ denotes a pair of nearest-neighbor sites, and $\square$ denotes an elementary plaquette. The first term in Eq.~\eqref{eq:HWilson} can be identified with the electric field energy, while the plaquette operator $U_\square$ measures the gauge invariant magnetic flux through a single plaquette, $\Phi \equiv \int d^2 \vec \sigma \cdot (\vec \nabla \times \vec A)$. Hence, the second term in Eq.~\eqref{eq:HWilson} is identified with the magnetic field energy.

In the lattice formulation of $U(1)$ gauge theories, the invariance of the underlying continuum theory under gauge transformations of the vector potential, $\vec A' = \vec A - \vec \nabla \alpha$, corresponds to an invariance of the Hamiltonian (\ref{eq:HWilson}) under lattice gauge transformations of the form 
\begin{equation}
\begin{split}
U'_{ij} &= V U_{ij} V^\dag = \exp({\rm i} \alpha_i) U_{ij}  \exp(- {\rm i} \alpha_j), \\
E'_{ij} &= V E_{ij} V^\dag = E_{ij}.
\end{split}
\end{equation}
Here $V = \prod_m \exp\{{\rm i} \alpha_m G_m\}$ is a unitary operator that implements a general gauge transformation.
Using the commutation relations between $E_{ij}$ and $U_{ij}$, one can convince oneself that the infinitesimal generator of
a gauge transformation at site $m$ is given by
\begin{equation}
G_{m}= E_{im} +  E_{km} - E_{mj} - E_{m\ell}.
\end{equation}
Note that $[H,G_m]=0$, so that the site charges $Q_m$, satisfying $G_m|\psi\rangle=Q_m|\psi\rangle$, are local conserved quantities under the time-evolution generated by $H$. In other words, for a specified charge configuration $\{Q_m\}$, the Gauss law $(G_m-Q_m)|\psi\rangle=0$ (for all $m$) defines a subset of physical states, where at each vertex the sum of incoming and outgoing fluxes is equal to the total charge at vertex $m$, $Q_m$. This condition is the lattice version of the usual Gauss law, $\vec{\nabla} \cdot \vec{E} = \rho$, for a continuous charge density $\rho$.  

The $U(1)$ QLM shares many features with the standard Wilson theory, but it uses a finite-dimensional representation of the local algebra $[E_{ij},U_{ij}]=U_{ij}$. This is possible because in QLMs the link variables $U_{ij}$ and $U^\dagger_{ij}$ are no longer complex numbers, but non-commuting operators. The quantum link operators obey $[U_{ij},U^\dagger_{ij}] = 2 E_{ij}$, which implies that $U_{ij}$, $U^\dagger_{ij}$, and $E_{ij}$ generate an $SU(2)$ embedding algebra on each link. $U(1)$ QLMs can be realized with any finite-dimensional spin $S$ representation of the $SU(2)$ algebra. In this case the electric flux on each link can only assume a finite set of discrete integer or half-integer values $e_{ij}$. The electric flux operator can then be identified with the third component of a spin $S$ operator, $S^{z}_{ij}$, and the quantum link variables are the corresponding raising and lowering operators, $S^{\pm}_{ij}$. More precisely, as illustrated in Fig~\ref{fig:general}(b)  for the case of $S=\frac{1}{2}$, the positive flux states around a single plaquette are mapped alternatingly into spin up and spin down states, according to $E_{ij} = S_{ij}^z$ and $U_{ij} = S_{ij}^-$ or $E_{ij} = - S_{ij}^z$ and $U_{ij} = S_{ij}^+$ [see Fig.~\ref{fig:general} for the mapping between fluxes and spins]. With this convention, the generators of the symmetry defined above are given by
\begin{equation}
G_m = S^z_{im} + S^z_{km} + S^z_{mj} + S^z_{m\ell},
\end{equation}
and the neutral subspace of the Hilbert space now corresponds to configurations with two spin up and two spin down states around each lattice site. 

The generators $G_m$ commute with the electric fluxes $E_{ij} = S_{ij}^z$ and with the plaquette operators $U_\Box = S_{ij}^+ S_{jk}^- S_{k\ell}^+ S_{\ell i}^-$. The spin $S$ representation of the Hamiltonian~\eqref{eq:HWilson} is then again invariant under $U(1)$ gauge transformations. A special scenario arises for the minimal $S=\frac{1}{2}$ representation, where the electric-field energy of equation \eqref{eq:HWilson} is $E_{ij}^2 = (S_{ij}^z)^2 = \frac{1}{4}$, and thus only contributes as a constant energy shift. In this case, a gauge invariant extension of the gauge field Hamiltonian can be considered, for example, of the form~\cite{BanerjeeU1,BanerjeeProceedings}
\begin{equation} 
\label{QLMham}
H = - J \sum_{\Box}\left[U_{\Box} + U^{\dagger}_{\Box}  - \lambda \left(U_{\Box} + U^{\dagger}_{\Box} \right)^{2}\right],
\end{equation}
where $\sum_\Box$ denotes the sum over all plaquettes. The first term (``kinetic energy'') inverts the direction of the electric flux around flippable plaquettes, while the second term (``potential energy'') favors the formation of flippable plaquettes. These terms are also known as ``ring-exchange'' and ``Rokhsar-Kivelson'' interactions, respectively. This Hamiltonian is gauge invariant, as it commutes with the generators of infinitesimal $U(1)$ gauge transformations $G_{m}$ given above.

The physics and phase diagram of this model is quite rich. At zero temperature, the model is confining for $\lambda<1$, while at high temperatures, $T>T_c$, it has a deconfining phase. At a critical coupling $\lambda_{c}$ there is a quantum phase transition, which separates two distinct confined phases with spontaneously broken translation symmetry. The phase at $\lambda < \lambda_{c}$ has, in addition, a spontaneously broken charge conjugation symmetry. The phase transition that separates the two confined phases is a weak first-order transition, but mimics several features of deconfined quantum critical points~\cite{BanerjeeU1}.

\subsection{Quantum dimer models}

In condensed matter physics, a closely related class of models are the so-called quantum dimer models. As we will see, they are also $U(1)$ gauge invariant, and describe the short-range resonating valence bond states proposed by Anderson \cite{AndersonRVB}, realizing valence bond solid or quantum spin liquid phases. Here, a dimer represents a singlet state formed by two electrons located at nearest-neighbor sites of a two-dimensional square lattice. Within the dimer model, the number of valence bonds is conserved, but they can rearrange themselves in such a way that each site shares exactly one dimer with one of the neighboring sites.

\begin{figure}[t]
\centering
\includegraphics[width=\columnwidth]{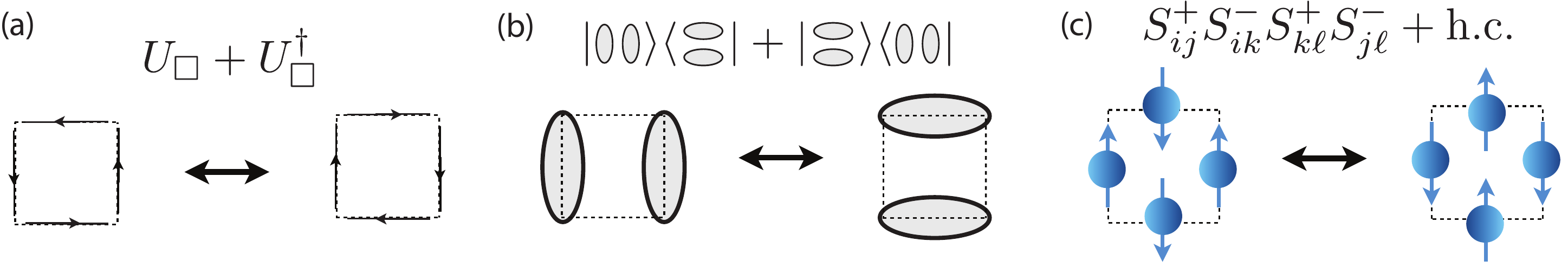}
\caption{ (Color online). Action of the ring-exchange Hamiltonian on flippable plaquettes. (a) Flow of electric flux through the links of the lattice. (b) Dimer covering. (c) Spin $\frac{1}{2}$ representation.}
\label{fluxdimerspin}
\end{figure}

In quantum dimer models, the degrees of freedom account for the presence or absence of a dimer on each link. According  to the dimer covering constraint, two dimers cannot touch each other, but can be located at opposite links of a lattice plaquette. The short-range dimer Hamiltonian can be written as \cite{RokhsarKivelson}
\begin{equation}
H_{\rm dimer} = - J \sum_{\Box} \Big[ \Big( \vertvbket \horvbbra + \horvbket \vertvbbra \Big) 
- \lambda \Big(  \vertvbket \vertvbbra + \horvbket \horvbbra \Big) \Big],
\end{equation}
where $\vertvbket$ and $\horvbket$ denote states with two dimers located vertically and horizontally, respectively, on opposite links of a plaquette. The relation between the dimer model and the spin $\frac{1}{2}$ QLM can be established by identifying the presence of a dimer with the state $e_{ij}= + \frac{1}{2}$ and the absence with the state $e_{ij} = - \frac{1}{2}$. With this identification, the Hamiltonian is recast into the form
\begin{equation} \label{DimerSpinEq}
\begin{split}
H_{\rm dimer} = - J\sum_{\Box} ( B_\square - \lambda B_\square^2),
\end{split}
\end{equation} 
where $B_\square = S_{ij}^{+} S_{ik}^{-} S_{k\ell}^{+} S_{j\ell}^{-} + {\rm H.c.}$, and which, using $U_\Box = S_{ij}^+ S_{jk}^- S_{k\ell}^+ S_{\ell i}^-$, corresponds to the quantum link model Hamiltonian \eqref{QLMham} [c.f. Fig.~\ref{fluxdimerspin} for the action of the ring-exchange interaction in lattice gauge theories, quantum dimer models, and quantum link models]. Although the $U(1)$ QLM and the dimer model share the same Hamiltonian, they differ in the realization of the Gauss law constraint, which for the dimer model is given by
\begin{equation}
Q_m = e_{im} +  e_{km} + e_{mj} + e_{m\ell} = - 1.
\end{equation}
This constraint ensures that exactly one dimer touches each lattice site. On the square lattice, around each site there are three links without a valence bond and just one link that carries a dimer. For $\lambda < 1$, the square lattice quantum dimer model exists in a confining columnar phase that extends to the Rokhsar-Kivelson point at $\lambda = 1$, a deconfined critical point at zero temperature.

\subsection{Confinement and string dynamics}

\begin{figure}[t]
\centering
\includegraphics[width=0.7\linewidth]{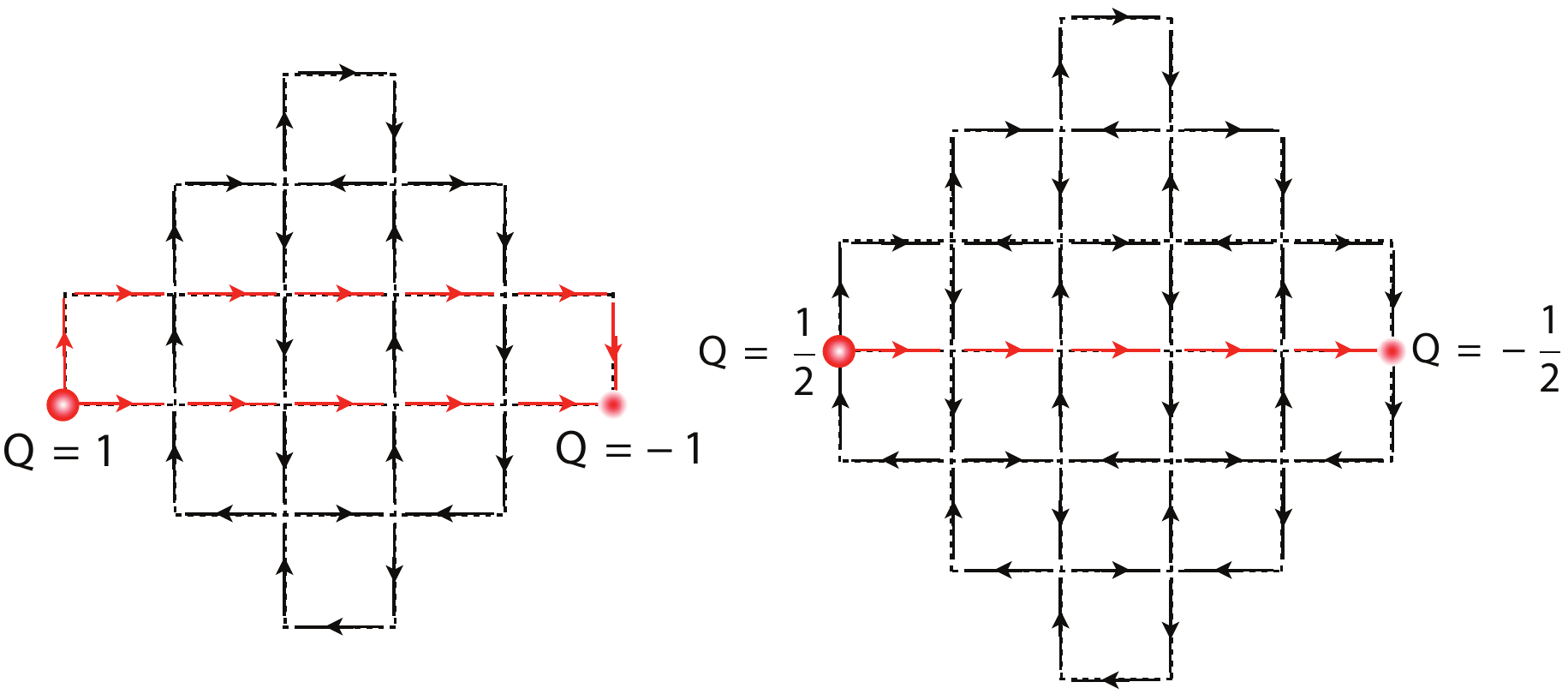}
\caption{(Color online). Illustration of possible strings of electric flux between a particle-antiparticle pair. Intrinsic properties of the string, such as its tension and width, contain fundamental information about confinement. Here we show two configurations with external charges $Q = \pm 1$ (left) and $Q = \pm \frac{1}{2}$ (right) at the boundaries. Flux strings connect the charge with the anticharge. The zig-zag boundary allows the Gauss law to be satisfied at the edges of the system.}
\label{string}
\end{figure}

\begin{figure}[t]
\centering
\includegraphics[width=0.45\columnwidth]{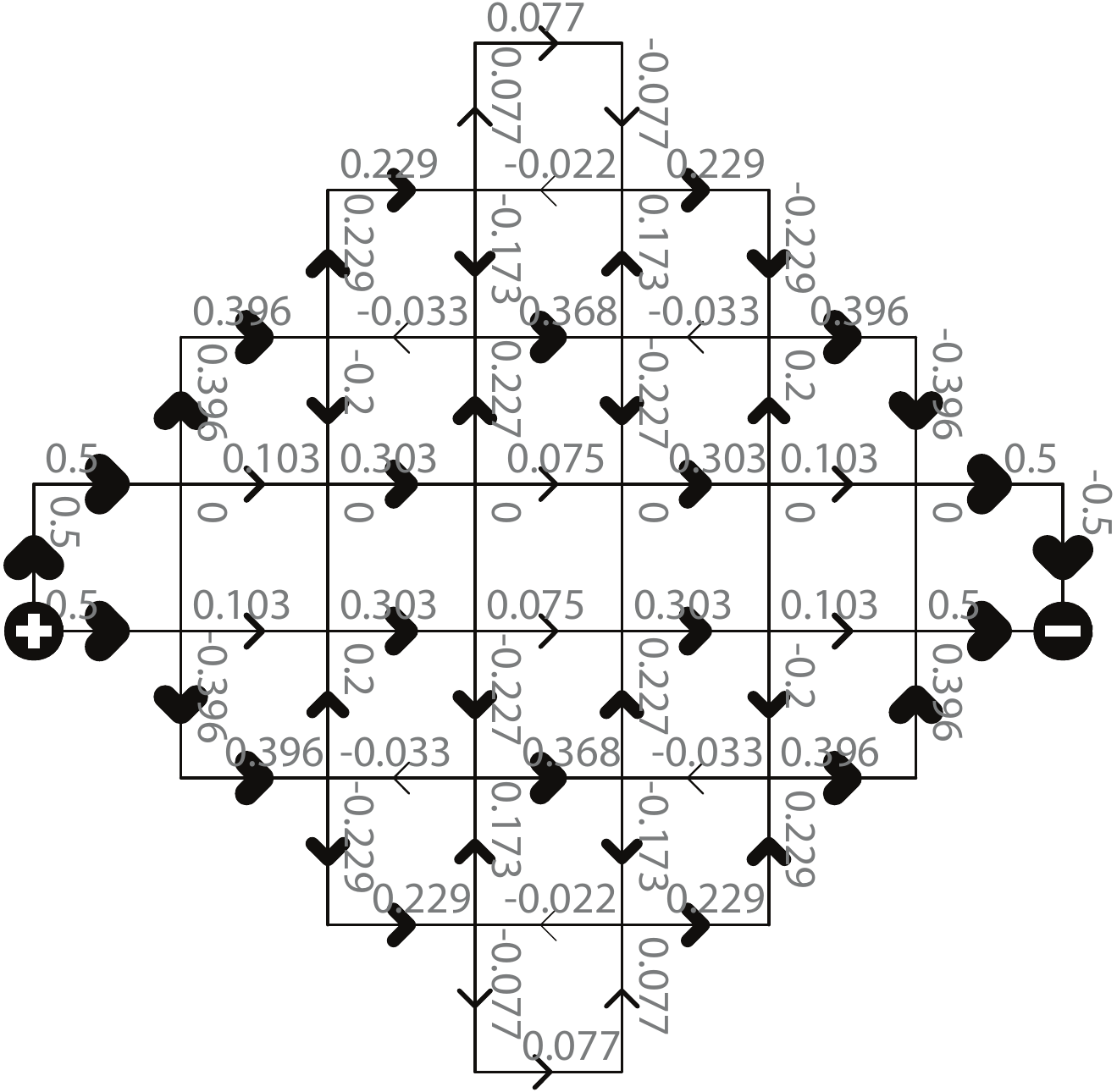}
\caption{Electric-field distribution for the ground state of the ring-exchange Hamiltonian \eqref{REhamiltonian} on a square lattice, using exact diagonalization. We have chosen open zig-zag boundaries in order to fulfil the Gauss law at each vertex. However, a charge-anticharge pair has been created at the edges by violating the Gauss law at those sites, giving rise to electric flux strings. The magnitude of the propagating electric flux is indicated on each link, and can be experimentally measured by taking snapshots of the spin distribution from an initially-prepared state.}
\label{stringdynamics}
\end{figure}

As mentioned above, the Gauss law, $G_m |\psi\rangle = 0$, can be violated by installing a charge-anticharge pair at two lattice sites. In this situation, the electric flux flows from particle to antiparticle [see Fig.~\ref{string} for illustrative examples and Fig.~\ref{stringdynamics} for an exact-diagonalization calculation], creating strings of flux whose tension and internal structure provide information about confinement: a string has an energy proportional to its length, with the string tension being the proportionality factor. In the two-dimensional $U(1)$ QLM a string connecting two particles of charge $Q_m=\pm 2$ separates into four mutually repelling strands, each carrying fractional electric flux $\frac{1}{2}$. Similarly, a string connecting particles of charge $Q_m =\pm 1$ splits into two strands. 

The excitation spectrum of the strings contains further physically relevant information. For example, it is interesting to see how the electric fluxes spread on the lattice in the transverse direction. This determines whether the strings separate into mutually repelling strands and whether they are rigid or rough. If the strings are rough, a continuum effective string theory describes their low-energy dynamics, which predicts that the width of the transverse string fluctuations grows logarithmically with the distance between the particle-antiparticle pair. The parameters of the effective string theory, such as the string tension and the intrinsic string width are measurable quantities. Below we present a roadmap for different experiments in small systems that begin to address these issues.

\subsection{Building blocks for simulating static and dynamical properties of quantum link models}

Given the broad interest in quantum link and quantum dimer models and their relevance in various areas of physics, in the remainder of this paper we address the controlled implementation of such models using coupled superconducting circuits. The main challenge in artificially engineering interactions of the type (\ref{DimerSpinEq}) is to realize the plaquette interactions between multiple spins. In this respect, superconducting circuits are potentially beneficial. First, different circuit elements can simply be connected via electrical wires. Second, the extremely large couplings and low losses observed in these systems allow the design of high-order interaction terms, which are sufficiently strong compared to the relevant decoherence energy scales. However, the fabrication and control of large arrays of superconducting qubits is still under development. Thus, it is the purpose of this work to first of all describe and analyze the implementation of the essential building blocks of QLMs, and to discuss the minimal settings which are required to observe precursors of the physical phenomena outlined above. This will provide a roadmap for constructing larger systems in a bottom-up approach. 

Of central interest to this work is the implementation of the ring-exchange plaquette interaction, $H_{\Box} = - J (U_{\Box} + U_{\Box}^{\dagger})$, which can be written in the spin notation as
\begin{equation} \label{REhamiltonian}
H_{\Box} = - J \left( S_{ij}^{+} S_{ik}^{-} S_{k\ell}^{+} S_{j\ell}^{-} + {\rm H.c.}\right).
\end{equation}
As already mentioned above, this interaction flips the spins around a plaquette and thus represents a delocalizing kinetic energy contribution. Besides the potential energy contribution $H_\Box^2$ of Eq.(\ref{QLMham}), we also consider a two-body spin interaction and first analyze the physics of the model
\begin{equation} \label{2Dmodel}
H = - J \sum_{\square} S_{ij}^{+} S_{ik}^{-} S_{k\ell}^{+} S_{j\ell}^{-} + {\rm H.c.} + V \sum_{\lefthalfcap} S_{ij}^z S_{jk}^z,
\end{equation}
where the last (gauge invariant) term represents an Ising-type coupling between adjacent link spins on each plaquette, which will be denoted by the symbol $\lefthalfcap$ in the sums, and favors spin configurations with a specific local magnetization. This model can be viewed as the simplest non-trivial extension of the pure ring-exchange interaction, and it exhibits a quantum phase transition as a function of $J/V$. Like in the QLM of Eq.(\ref{QLMham}), the transition separates two distinct confined phases. 

A more general gauge invariant model for spin $\frac{1}{2}$ is given by the Hamiltonian
\beq \label{2Dmodel2}
H = - J \sum_{\square} \left(S_{ij}^{+} S_{ik}^{-} S_{k\ell}^{+} S_{j\ell}^{-} + {\rm H.c.} \right) + V \sum_{\lefthalfcap} S_{ij}^z S_{jk}^z 
+ W \sum_{\square} S_{ij}^{z} S_{ik}^{z} S_{k\ell}^{z} S_{j\ell}^{z}.
\eeq
Here, in addition to the two-body interaction, we have included a four-body plaquette term that favors an odd number of spins pointing along the same direction around every plaquette. The combination of ring-exchange, two-body nearest-neighbor interaction, and four-body plaquette interaction, gives a large class of models that, as we show below, can be quantum simulated with superconducting circuits. Next, we show the corresponding implementation, and how the associated nontrivial dynamics can be probed in experiments. 

\section{Superconducting circuit implementation} \label{SCimplementationSec}

The ring-exchange Hamiltonian \eqref{REhamiltonian} involves non-local four-body interactions, which do not appear naturally in superconducting circuits or systems with dipolar interactions. In the following, we describe how this type of interactions can be implemented using quantized excitations in electrical circuits. As a concrete example, we will focus on a circuit layout based on `transmon' qubits \cite{KochTransmon}; however, the scheme is quite general and can be adapted to other superconducting-qubit implementations as well.

\subsection{General approach}

\begin{figure}[t]
\begin{center}
\includegraphics[width=0.4\linewidth]{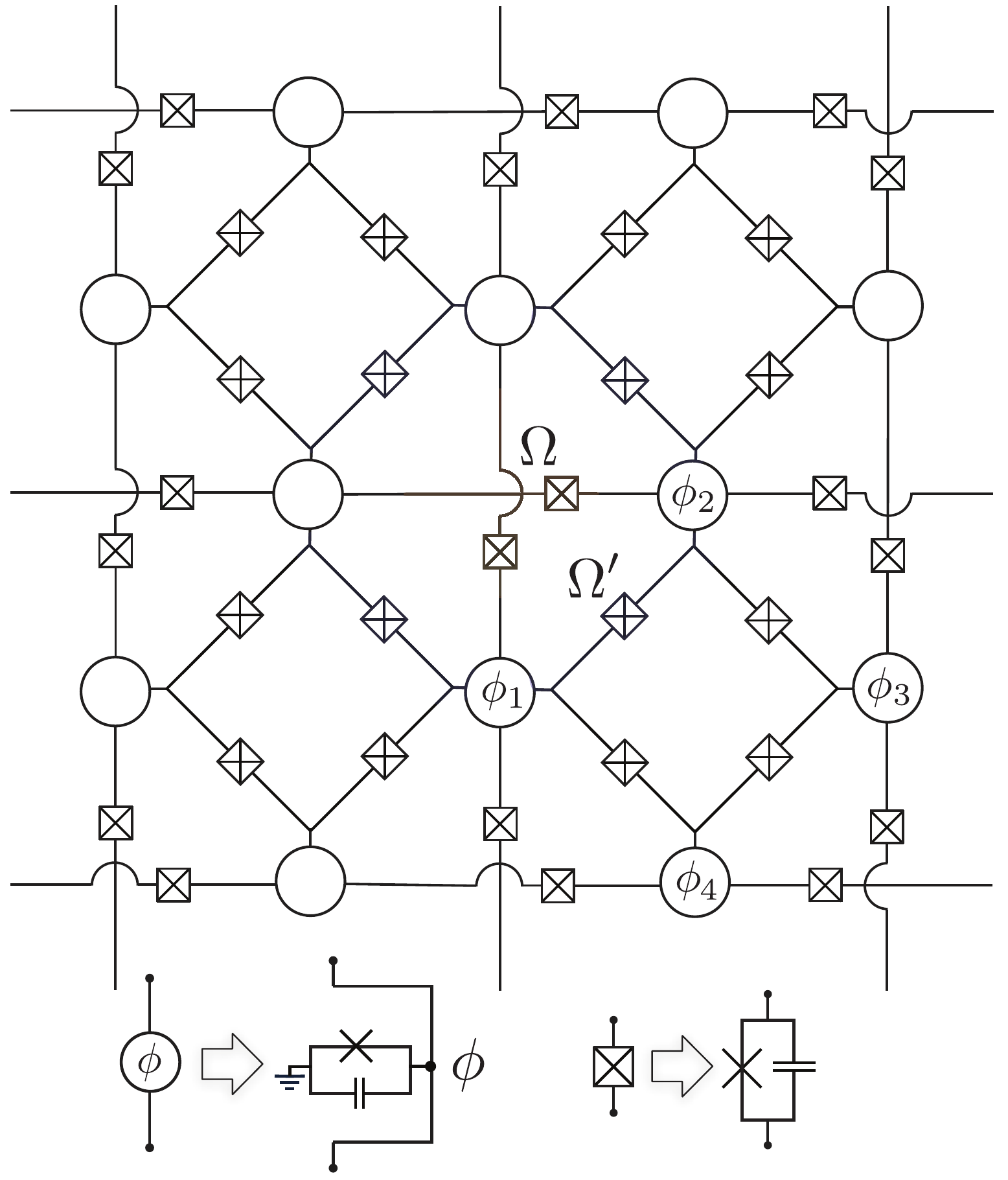}
\caption{Circuit lattice for the simulation of the model (\ref{2Dmodel}). Every plaquette contains a qubit (e.g.\ a transmon) on the links. Hopping and Kerr interactions of local excitations are enabled by a capacitor in parallel with a Josephson junction connecting neighboring qubits, giving rise to $S_{ij}^z S_{jk}^z$ interactions, and --- perturbatively --- to ring-exchange dynamics. The tunneling term through each vertex may be suppressed by choosing appropriately the value of the parallel capacitor to the Josephson junction.}
\label{latticeFig}
\end{center}
\end{figure}

Let us consider the general circuit lattice depicted in Fig.~\ref{latticeFig}. On each link the lowest two energy levels of a strongly coupled superconducting circuit (qubit) are used to implement an effective spin $\frac{1}{2}$ system, representing the gauge field, as described in Sec.~\ref{QLMsec}.
Neighboring spins on each plaquette and across each node are connected by Josephson junctions, which induce nearest-neighbor interactions. By an appropriate choice of parameters, the resulting Hamiltonian of the circuit lattice takes the form 
\beq \label{MotherH}
H = \varepsilon \sum_{\langle ij \rangle} S_{ij}^z - \Omega  \sum_{+} S_{ij}^z S_{jk}^z - \Omega' \sum_{\lefthalfcap} S_{ij}^z S_{jk}^z -  \mu \sum_{\lefthalfcap} ( S_{ij}^{+} S_{jk}^{-} + {\rm H.c.} ),
\eeq
where $\varepsilon$ is the bare frequency splitting between qubit states (the sum $\sum_{\langle ij \rangle}$ involves nearest-neighbor lattice sites). The interactions $\Omega$ and $\Omega'$ are diagonal coupling constants for qubits located on opposite sides of each lattice site and neighboring qubits within the same plaquette, respectively [see Fig.~\ref{latticeFig}] (the sum $\sum_+$ denotes qubits around vertices, and the sum $\sum_{\lefthalfcap}$ involves nearest-neighbor links around a plaquette). In addition, neighboring qubits located within the same plaquette are coupled by a small hopping term $\sim \mu$. By defining $V'=\Omega-\Omega'$ and omitting an overall frequency shift, we can rewrite the Hamiltonian \eqref{MotherH} as
\beq \label{HmicroG}
H = \varepsilon \sum_{\langle ij \rangle} S_{ij}^z - \Omega \sum_m G_m^2  + V'\sum_{\lefthalfcap} S_{ij}^z S_{jk}^z - \mu \sum_{\lefthalfcap} ( S_{ij}^{+} S_{jk}^{-} + {\rm H.c.} ),
\eeq
where for each site $G_m=S^z_{im} +  S^z_{km} + S^z_{mj} + S^z_{m\ell}$ is the gauge generator introduced above. Under the assumption that the system is initially prepared in the subspace of states with exactly two spins up and two spins down around each site, $G_m|\psi\rangle=0$ for all $m$, transitions out of this subspace are suppressed by a large energy gap $\Omega$. In the limit $\mu, V' \ll \Omega$ we can use perturbation theory to derive an effective Hamiltonian for this subspace, which is given by 
\beq \label{HGringEff}
H_{\rm eff} = \varepsilon \sum_{\langle ij \rangle} S_{ij}^z + V \sum_{\lefthalfcap} S_{ij}^z S_{jk}^z - J \sum_{\square} (S_{ij}^{+} S_{ik}^{-} S_{k\ell}^{+} S_{j\ell}^{-} + {\rm H.c.}),
\eeq
where
\begin{equation} \label{JVparam}
J= \frac{4\mu^2}{\Omega}, \qquad V= V' - J.
\end{equation}
Apart from the overall qubit energy $\sim \varepsilon$, which does not affect the dynamics in the gauge invariant subspace, this effective Hamiltonian reproduces the gauge invariant model (\ref{2Dmodel}). In particular, taking $V=0$, the standard ring-exchange interaction \eqref{REhamiltonian} is recovered. An interaction of the type $S_{ij}^{z} S_{ik}^{z} S_{k\ell}^{z} S_{j\ell}^{z}$ (arising in the RK model) requires an additional circuit element, which will be discussed in Sec.~\ref{RKsec}.

\subsection{Circuit model}

We now show how the aforementioned interactions can be implemented using superconducting circuits, in particular using transmon qubits on the links of a two-dimensional lattice [c.f. Fig.~\ref{latticeFig}]. A single transmon consists of a capacitance $C$ in parallel with a Josephson junction with energy $E_J$. This circuit is described by a Hamiltonian
\begin{equation} \label{transmonH}
H_{\rm transmon} =  \frac{Q^2}{2C}- E_J  \cos\left( \frac{\phi}{\phi_0}\right),
\end{equation}
where $Q$ and $\phi$ are the canonically conjugate charge and flux operators, obeying $[\phi,Q]=i$, and $\phi_0=1/(2e)$ is the reduced flux quantum ($\phi_0\approx 0.33\times 10^{-15}$ Wb) [here we take $\hbar \equiv 1$]. In the regime where the Josephson energy $E_J$ dominates over the charging energy $E_C=e^2/(2C)$, the flux fluctuations are small compared to $\phi_0$, and the cosine potential in Eq.~(\ref{transmonH}) can be expanded in powers of $\phi/\phi_0$. Up to fourth order in this expansion, we then obtain the Hamiltonian of a non-linear oscillator \cite{KochTransmon}
\begin{equation}
H_{\rm transmon} \approx \frac{Q^2}{2C} + E_J\frac{\phi^2}{2\phi_0^2} - E_J \frac{\phi^4}{24\phi_0^4}\approx \varepsilon a^\dag a- \frac{U}{2} a^\dag a^\dag a a,
\end{equation}
where we have introduced annihiliation and creation operators $a$ and $a^\dag$ according to 
\begin{equation}
 \frac{Q}{2e} =\sqrt[4]{\frac{E_J}{8E_C}} \, \frac{i(a^\dag- a)}{\sqrt{2}}, \;\;\;\;  \frac{\phi}{\phi_0}=  \sqrt[4]{\frac{8E_C}{E_J}}\, \frac{(a+a^\dag)}{\sqrt{2}}.
\end{equation} 
For typical experimental parameters, the qubit frequency $\varepsilon=\sqrt{8E_CE_J} - U$ is several GHz, and the strength of the nonlinearity $U\approx E_C$ is around several $100$ MHz \cite{Schreier}. Assuming that this nonlinearity is sufficiently large to prevent transitions into states with $n\geq2$ excitations, the dynamics of the transmon can be restricted to the lowest two oscillator states, $|\downarrow\rangle \equiv |0\rangle$ and $|\uparrow\rangle\equiv |1\rangle$, and modeled by a spin $\frac{1}{2}$ Hamiltonian,
\begin{equation}
H_{\rm transmon} \approx \varepsilon  S^z. 
\end{equation} 

\begin{figure}[t]
\centering
\includegraphics[width=0.55\linewidth]{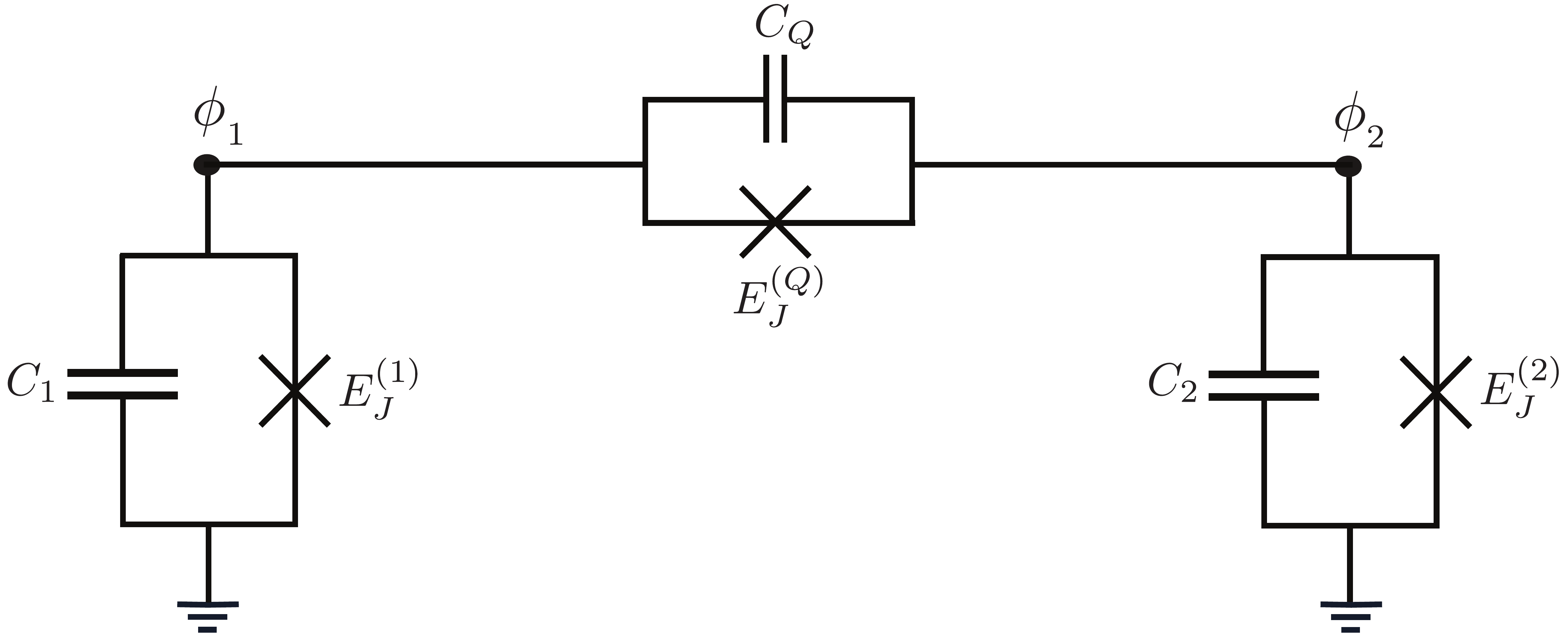}
\caption{Basic building block for the lattice gauge theory architecture shown in Fig.~\ref{latticeFig}. Two superconducting qubits (transmons) are coupled through a Josephson junction in parallel with a capacitor. This enables hopping and Kerr interactions between quantized photonic excitations at nodes $1$ and $2$. The value of the capacitor can be chosen appropriately in order to control the hopping of excitations.}
\label{transmons_coupled}
\end{figure}

To implement interactions between neighboring qubits, we now consider the basic building block shown in Fig.~\ref{transmons_coupled}. Here two transmons are connected via an additional Josephson junction with energy $E_J^{(Q)}$ in parallel with a capacitor $C_Q$.
The associated Hamiltonian is \cite{Marcos13, Tian, Hartmann13}
\beq \label{Hcircuit}
H = \frac{1}{2} \vec{Q} {\cal C}^{-1} \vec{Q}^T  - \sum_{\ell = 1,2} E_J^{(\ell)} \cos \left( \frac{\phi_{\ell}}{\phi_0} \right) -  E_J^{(Q)} \cos \left( \frac{\phi_{1}-\phi_{2}}{\phi_0} \right),
\eeq
where $Q_\ell$ and $\phi_\ell$ are the charge and flux operators at  a node $\ell$, $\vec{Q} \equiv (Q_1,Q_2)$, and 
\begin{equation}
 \mathcal{C}=
\left(\begin{array}{cc}
C_1 +C_Q  & - C_Q \\
-C_Q & C_2 +C_Q \\
\end{array}\right),
\end{equation}
is the capacitance matrix. As above, for small phase fluctuations we can expand the cosine functions and write the resulting Hamiltonian as
\begin{equation} \label{Hbuildingblock}
H = \sum_{\ell=1,2} H_\ell + H_{\rm int}. 
\end{equation}   
Here, 
\begin{equation}
H_\ell = \frac{Q_{\ell}^2}{2\bar C_\ell} + \left(E^{(\ell)}_J+E_J^{(Q)}\right) \frac{\phi_\ell^2}{2\phi_0^2}-\left(E^{(\ell)}_J+E_J^{(Q)}\right)  \frac{\phi_\ell^4}{24\phi_0^4}, \quad\quad
\end{equation} 
are the modified Hamiltonians for each qubit, where 
\begin{equation}
\bar C_1= C_1 +\frac{C_2C_Q}{C_2+C_Q},\qquad \bar C_2= C_2 +\frac{C_1C_Q}{C_1+C_Q}.
\end{equation} 
By assuming that $C_Q < C_\ell$ and $E_J^{(Q)} <  E_J^{(\ell)}$, the coupling junction does not qualitatively change the single-qubit Hamiltonians, $H_\ell\approx \varepsilon_\ell S_\ell^z$, with slightly modified frequencies $\varepsilon_\ell$.
The remaining interaction Hamiltonian is given by
\begin{equation}
H_{\rm int}\approx \frac{C_Q}{C_1 C_2}  Q_1Q_2 - \frac{E_J^{(Q)}}{\phi_0^2} \phi_1\phi_2 - \frac{E_J^{(Q)}}{4\phi_0^4} \phi_1^2\phi_2^2+ \frac{E_J^{(Q)}}{6\phi_0^4} \left( \phi_1\phi_2^3+\phi_1^3\phi_2\right),
\end{equation}
and when projected onto the spin subspace of interest, we obtain
\begin{equation}
H_{\rm int}\approx - \frac{\Omega}{2}(S_1^z+S_2^z) - \mu (S_1^+S_2^-+S_1^-S_2^+) - \Omega S_1^z S_2^z.
\end{equation} 
We notice that here the subindexes $1$ and $2$ refer to respective circuit nodes of Fig.~\ref{transmons_coupled}, which are located on the links of the two dimensional lattice of Fig.~\ref{latticeFig}. The first term in this Hamiltonian is a small frequency shift, which can be absorbed into a redefinition of the qubit frequency, $\varepsilon_\ell \rightarrow \varepsilon_\ell -\Omega/2$. The other two contributions represent a spin flip-flop and an Ising-type spin-spin interaction with  coupling strengths 
\beq
\mu = \frac{\varepsilon}{2} \left(\frac{E_J^{(Q)}}{E_J}  - \frac{C_Q}{C}\right)  -\Omega, \quad 
\Omega = U \frac{2E_J^{(Q)}}{E_J}, \label{OmegaEq}
\eeq
where we have assumed $E_J^{(1)} = E_J^{(2)} \equiv E_J$. Still under the assumption that the capacitance $C_Q$ and the Josephson energy $E_J^{(Q)}$ are sufficiently small, the coupling between different neighboring transmons on the lattice of Fig.~\ref{latticeFig} can simply be added up. Considering different coupling constants around plaquettes ($\square$) and across lattice sites ($+$), and taking
\begin{equation}
\begin{split}
\mu_{\square} &= \mu, \quad \Omega_{\square} = \Omega', \\
\mu_{+} &= 0, \quad \Omega_{+} = \Omega,
\end{split}
\end{equation}
we obtain the model (\ref{MotherH}), from which we then derive the effective Hamiltonian (\ref{HGringEff}), with parameters $J$ and $V$ as defined in Eq.~(\ref{JVparam}).

\subsection{Parameters and tunablity}

\begin{figure*}[t]
\centering
\includegraphics[width=\linewidth]{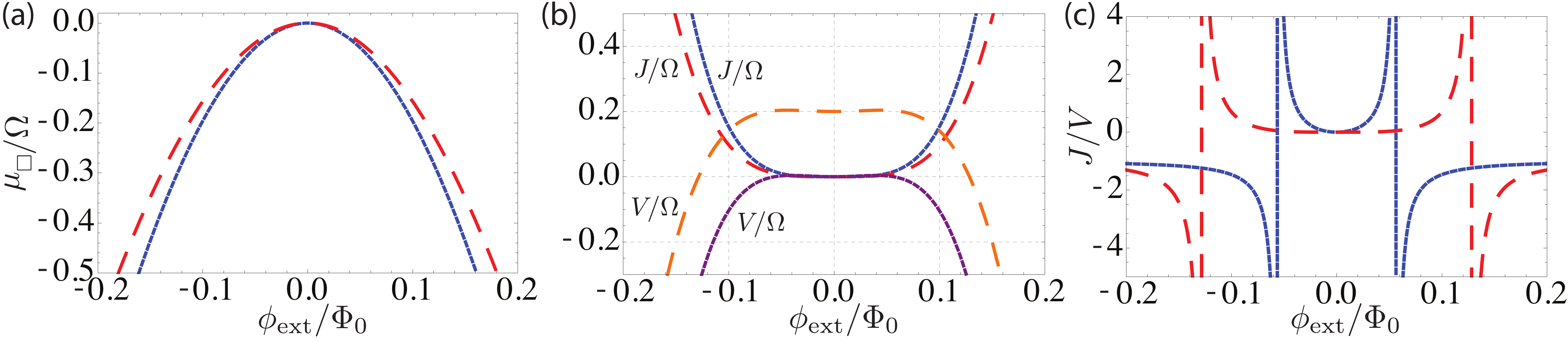}
\caption{(Color online). Different parameter values as a function of the external flux $\phi_{\rm ext}$. Here $\varepsilon = 6$ GHz, $U = 300$ MHz, $C_+/C = C_\square/C = 0.16$, $E_J^+/E_J = E_J^\square/E_J = 0.2$ (dotted solid lines), and $C_+/C = 0.20$, $C_\square/C = 0.16$, $E_J^+/E_J = 0.25$, $E_J^\square/E_J = 0.20$ (dashed lines). (a) The ratio $\mu_\square/\Omega$ determines the region of external flux in which perturbation theory is still valid. (b) Behavior of $J/\Omega$ and $V/\Omega$, [$\Omega = 120$ MHz (dotted solid lines) and $\Omega = 150$ MHz (dashed lines)]. Tuning the external magnetic flux, the regimes i) $J=0$, $V\neq 0$, ii) $J=V\neq 0$, and iii) $J\neq0$, $V=0$, can be reached. (c) Tunability of the ratio $J/V$. For the situation plotted with dashed lines, at $\phi_{\rm ext}/\Phi_0 \approx 0.13$ we find $V\approx 0$, and $J\neq 0$, giving rise to a ring-exchange interaction only. In the vicinity of that point, the ratio $J/V$ can go from positive to negative values.
}
\label{parameters}
\end{figure*}

For the simulation of the model \eqref{2Dmodel} we require that the effective parameters $J$ and $V$ are much larger than the relevant decoherence rates of the qubits, and that the ratio $J/V$ is tuneable to explore different regimes.
Assuming that the capacitances are fixed, the relative strength of the model parameters can be adjusted by tuning the Josephson energies $E_J^{(Q)}$. This can be done by replacing a single junction by an equivalent two-junction loop, with an effective Josephson energy given by
\beq \label{EJsubstitution}
E_J^{(Q)} \to E_J^{(Q)} \cos \left( \pi \frac{\phi_{\rm ext}}{\Phi_0} \right),
\eeq
where $\phi_{\rm ext}$ denotes an external magnetic flux through the loop, and $\Phi_0 \equiv 2\pi \phi_0$ is the magnetic flux quantum. We then set $C_Q = C_+$, $E_J^{(Q)}=E_J^{+}$, and $C_Q=C_\square$, $E_J^{(Q)}=E_J^{\square}\cos \left( \pi \phi_{\rm ext}/\Phi_0\right)$, for the couplings across the lattice sites and within each plaquette, respectively (thus making the latter tuneable), and choose the circuit parameters such that
\begin{equation} \label{eq:ZeroTunnelingPlus}
\mu_+ = \frac{\varepsilon}{2} \left(\frac{E_J^{+}}{E_J}  - \frac{C_+}{C}\right)  -\Omega=0,
\end{equation} 
and
\begin{equation}\label{eq:ZeroTunnelingSquare}
\mu_\square = \frac{\varepsilon}{2} \left(\frac{E_J^{\square}}{E_J}  - \frac{C_\square}{C}\right)  -\Omega'=0.
\end{equation}
The coupling constants $\mu_\square$ and $\Omega_\square$, and therefore the ratio $J/V$, can now be tuned by considering a two-junction loop, coupling neighboring links around the plaquette. These loops can be biased using either a global magnetic field or local flux lines to generate a finite $\phi_{\rm ext}$ for the $\square$-links, replacing the value of $E_J^{\square}$ and $\Omega'$ in Eq.~\eqref{eq:ZeroTunnelingSquare} by $E_J^{\square} \cos \left( \pi \frac{\phi_{\rm ext}}{\Phi_0} \right)$ and $\Omega' \cos \left( \pi \frac{\phi_{\rm ext}}{\Phi_0} \right)$, respectively. This generates a finite $\mu_\square\neq 0$, which increases $J$ and simultaneously lowers $V$. At a certain value of the external flux, we reach $V=0$, and we recover the pure ring-exchange interaction. When $\phi_{\rm ext}=0$, we have $J=0$ and $V = \frac{2U}{E_J} (E_J^+ - E_J^\square)$.

In Fig.\ \ref{parameters} we show the behavior of the different system parameters as a function of the external flux. A fine-tuning of the $C_Q$'s ensures that for $\phi_{\rm ext}$ the conditions (\ref{eq:ZeroTunnelingPlus}) and (\ref{eq:ZeroTunnelingSquare}) are fulfilled. Typical values of the coupling constants in the region of magnetic flux where the perturbative approach leading to Eq.~\eqref{HGringEff} is still valid ($\mu_\square/\Omega \lesssim 0.5$) are $\Omega \sim 50$ MHz, $\mu, J, V \sim 5$ MHz, still much larger than the standard decoherence rates of a few tens of kHz. As we will show below, the tunability shown in Fig.~\ref{parameters} allows us to access the different phases of the model (\ref{HGringEff}).

\subsection{Rokhsar-Kivelson model} \label{RKsec}

\begin{figure}[t]
\begin{center}
\includegraphics[width=0.4\linewidth]{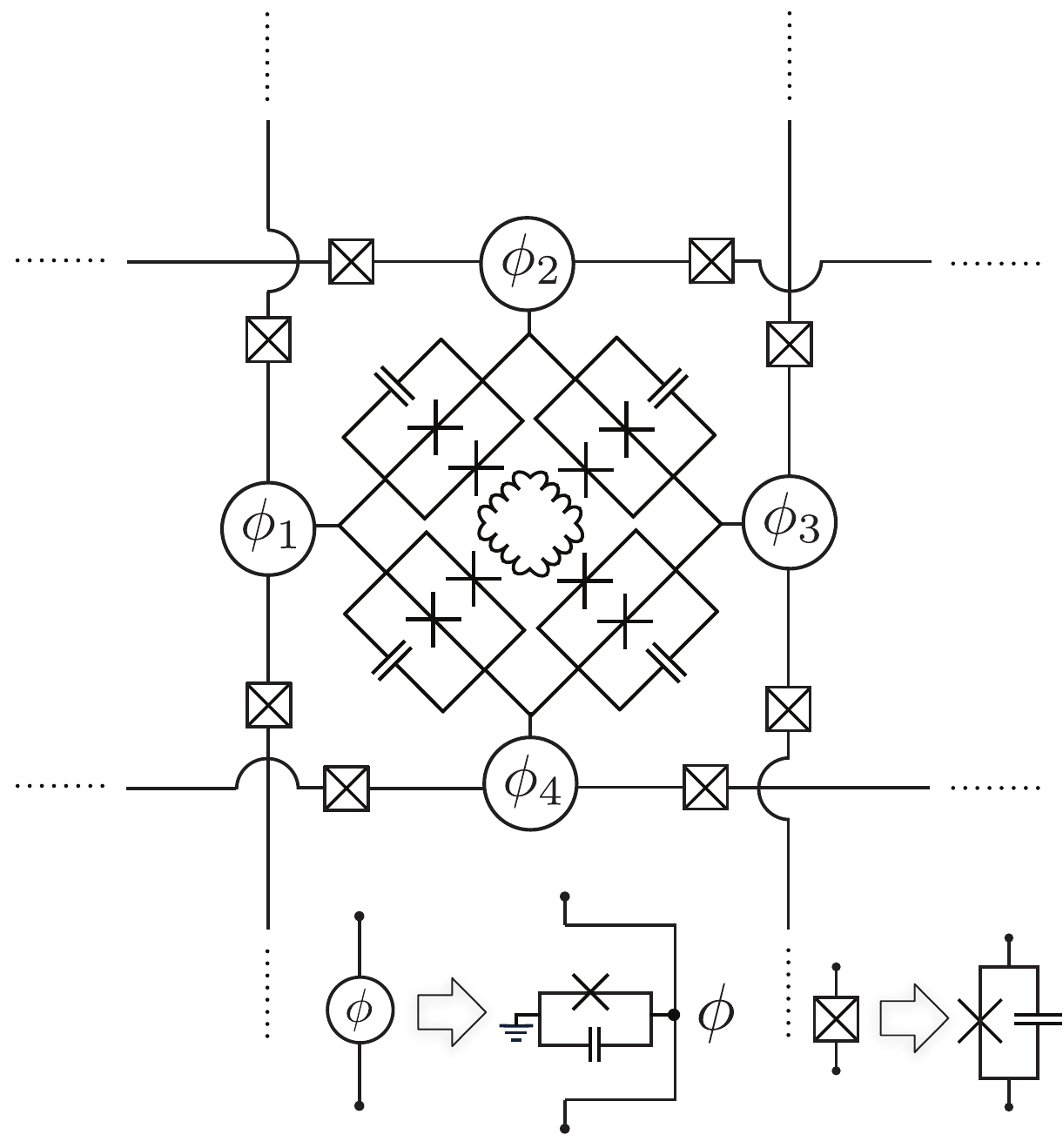}
\caption{Circuit lattice to engineer the Rokhsar-Kivelson model and different four-body spin interactions. Every plaquette of the two-dimensional lattice contains one qubit (e.g.\ a transmon) on each link. These are mutually coupled via a capacitor in parallel with a two-Josephson-junction loop. When this loop is biased with a quantum flux from a central LC circuit, interactions of the type $\sim S_1^zS_2^zS_3^zS_4^z$ are enabled perturbatively (see main text for details).}
\label{FigRK}
\end{center}
\end{figure}

Different gauge invariant interactions can be engineered by slightly modifying the complexity of the circuit lattice shown in Fig.~\ref{latticeFig}. A particularly interesting example is the Rokhsar-Kivelson (RK) model \cite{RokhsarKivelson} --- a paradigm of dimer physics, which describes resonant valence bond dynamics, relevant in the context of high-temperature superconductivity \cite{AndersonRVB}. This model can be simulated with the circuit shown in Fig.~\ref{FigRK}, where we draw a basic plaquette of the two-dimensinal lattice. Although this circuit is similar to the architecture of Fig.~\ref{latticeFig}, here the squids coupling neighboring transmons are biased with a {\it quantum flux} from an LC resonator located at the center of the plaquette. Following a similar derivation to section \ref{SCimplementationSec}, the model describing this circuit can be written as
\begin{equation} \label{HmicroRK}
\begin{split}
H &= \omega b^{\dagger} b + \varepsilon \sum_{\langle ij \rangle} S_{ij}^z - \Omega  \sum_m G_m^2  + V' \sum_{\lefthalfcap} S_{ij}^z S_{jk}^z - \mu \sum_{\lefthalfcap} ( S_{ij}^{+} S_{jk}^{-} + {\rm H.c.} ) \\ &+ (b^{\dagger} + b) \; \Big[ \beta' \sum_{\lefthalfcap} \varsigma_{ij} S_{ij}^z S_{jk}^z - \eta \sum_{\lefthalfcap} \varsigma_{ij} (S_{ij}^+ S_{jk}^- + {\rm H.c.}) \Big].
\end{split}
\end{equation}
Here $\varsigma_{ij}=1$ for spins on horizontal links of the lattice, while $\varsigma_{ij}=-1$ for vertical links. The sum $\sum_{\langle ij \rangle}$ involves nearest-neighbor lattice sites, and the sum $\sum_{\lefthalfcap}$ involves nearest-neighbor links around a plaquette.
For equal transmons, and in the limit $C_Q \ll C_\ell$, $E_J^{(Q)} \ll E_J^{(\ell)}$, the coupling constants are given by
\begin{equation}
\begin{split}
&V' = \Omega - \Omega', \quad
\Omega' = U \frac{2E_J^\square}{E_J} \cos \left( \pi \frac{\phi_{\rm ext}}{\Phi_0} \right), \quad
\mu = \frac{\varepsilon}{2} \left(\frac{E_J^\square}{E_J} \cos \left( \pi \frac{\phi_{\rm ext}}{\Phi_0} \right)  - \frac{C_Q}{C}\right)  -\Omega', \\ 
&\beta' = U \frac{2E_J^\square}{E_J} \sin \left( \pi \frac{\phi_{\rm ext}}{\Phi_0} \right), \quad
\eta = \frac{\varepsilon}{2} \frac{E_J^\square}{E_J} \sin \left( \pi \frac{\phi_{\rm ext}}{\Phi_0} \right) - \beta'.
\end{split}
\end{equation}
In the derivation of the Hamiltonian (\ref{HmicroRK}) we have assumed that, on top of the quantum flux from the resonator, consecutive squids are biased with external classical fields of alternating signs. Furthermore, we notice that, under realistic experimental conditions, the constants $\beta'$ and $\eta$ will be reduced by a factor $\alpha\leqslant 1$ determined by the fraction of the LC-resonator flux biasing the squid.

Given the Hamiltonian (\ref{HmicroRK}), and the hierarchy of scales $V', \mu, \beta', \eta \ll \Omega \ll \varepsilon, \omega$, we can treat the terms $\sim V', \mu, \beta', \eta$ perturbatively, and obtain the second-order effective dynamics
\beq \label{RKnearly}
H_{\rm eff} =  \omega b^{\dagger} b + \varepsilon \sum_{\langle ij \rangle} S_{ij}^z - \Omega  \sum_m G_m^2
- J \sum_{\square} B_\square + V \sum_{\square} B_\square^2 - \frac{V}{2} \sum_{||} S_{ij}^z S_{k\ell}^z.
\eeq
Here $\sum_{||}$ restricts the sum to opposite links on each plaquette, $B_\square \equiv S_{ij}^{+}S_{kj}^{-}S_{k\ell}^{+}S_{i\ell}^{-} + {\rm H.c.}$, and the coupling constants are given by $J = - \frac{4\mu^2}{\Omega} - \frac{4\eta^2}{\Omega-\omega}$, $V = - \frac{2\beta^2}{\omega}$. Furthermore, we have taken parameters such that $V' = -J - V/2$, and assumed that the central resonator is initially cooled to the ground state, thus having transitions between the resonator Fock states $|0\rangle$ and $|1\rangle$ only. The last term of Eq.~(\ref{RKnearly}) can be eliminated by adding a Josephson junction in parallel with a capacitor connecting opposite links on each plaquette, and choosing the corresponding capacitance and Josephson energies appropriately. In this case, using $B_\square = U_\square + U_\square^\dagger$ and identifying $\lambda = V/J$, the Hamiltonian (\ref{RKnearly}) reproduces the dynamics given by (\ref{QLMham}). Alternatively, choosing $V'=-J$, we obtain the effective Hamiltonian
\beq
H_{\rm eff} =  \omega b^{\dagger} b + \varepsilon \sum_{\langle ij \rangle} S_{ij}^z - \Omega  \sum_m G_m^2 - J \sum_{\square} (S_{ij}^{+} S_{kj}^{-} S_{k\ell}^{+} S_{i\ell}^{-} + {\rm H.c.}) + 2V \sum_{\square} S_{ij}^z S_{kj}^z S_{k\ell}^z S_{i\ell}^z,
\eeq
which displays explicitly the competition between ring-exchange and a four-body Ising interaction.

\section{Probing Ring-Exchange Interactions} \label{1plaquetteSec}

\begin{figure}
\centering
\includegraphics[width=0.55\linewidth]{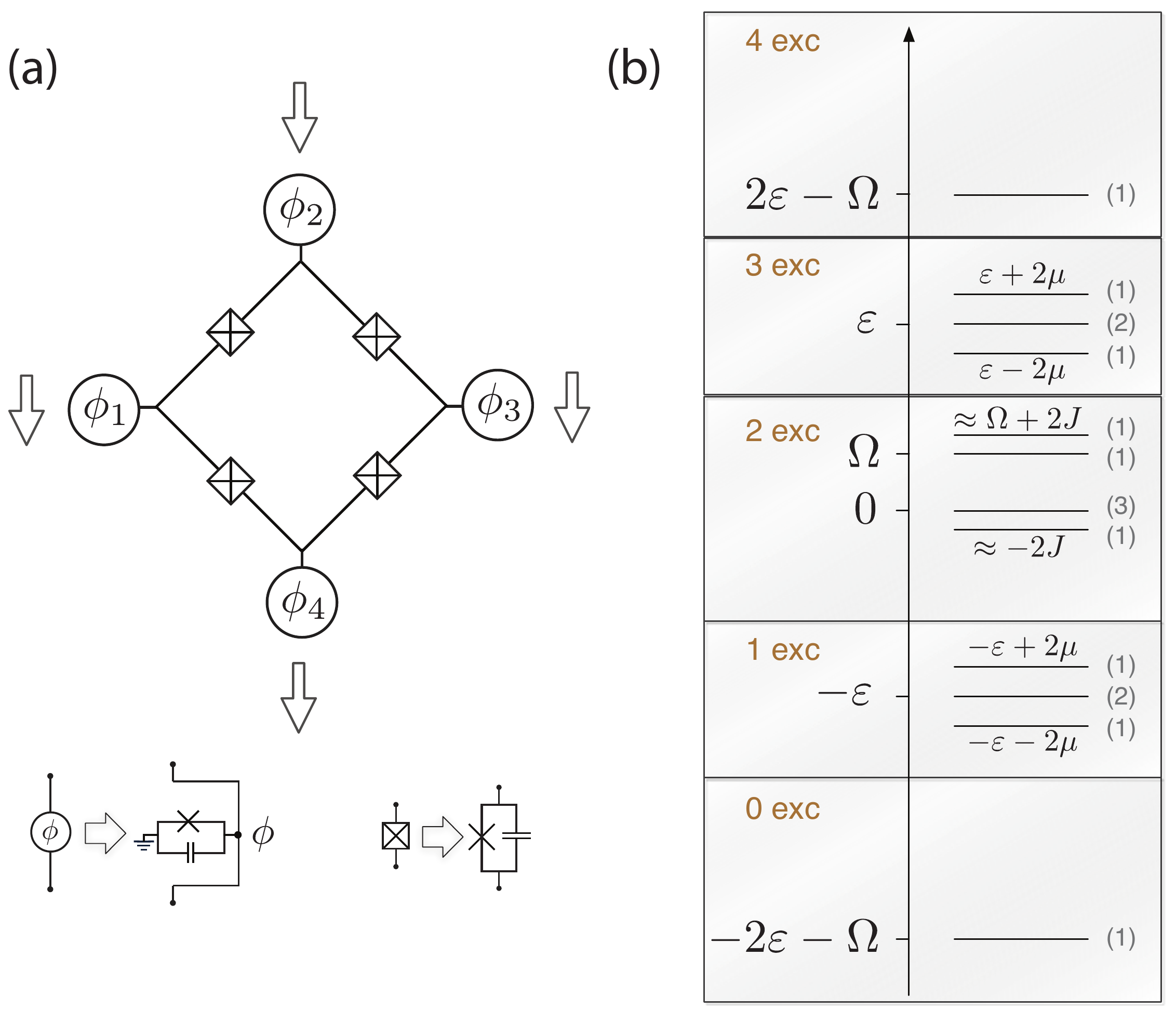}
\caption{(Color online). (a) Setup for a minimal experiment to verify ring-exchange dynamics. This system consists of a single plaquette on the lattice of Fig.~\ref{latticeFig}, where four superconducting qubits (whose spin degree of freedom is represented by the arrows) are mutually coupled via a Josephson junction in parallel with a capacitor. (b) Energy levels of the microscopic Hamiltonian (\ref{MotherH}) [see~\ref{eigenstates_app}]. Five different sets of states (distinguished by the total number of excitations on the plaquette) are separated by the energy scale $\varepsilon$ (qubit frequency). Within the two-excitation subspace, a large energy scale $\Omega$ separates states corresponding to different Gauss law sectors. Finally, the lower energy scales $\mu$ and $J$ provide an energy splitting (in the one- and two-excitation subspaces, respectively). The numbers on the right indicate the level degeneracy.}\label{plaquetteRing}
\end{figure}

A minimal setup for studying ring-exchange interactions is a circuit with four superconducting qubits forming a single plaquette [see Fig.~\ref{plaquetteRing}]. 
The approach described in the previous section can then be used to engineer an effective ring-exchange interaction within the two-excitation subspace of the four spins on the plaquette. In this minimal instance, the only non-vanishing coupling is between the states $|\uparrow \downarrow \uparrow \downarrow\rangle$ and $|\downarrow \uparrow \downarrow \uparrow\rangle$, i.e.,
\begin{equation}
 \langle \downarrow \uparrow \downarrow \uparrow | H | \uparrow \downarrow \uparrow \downarrow\rangle = -J.
\end{equation} 
Note that for a single plaquette the Ising-type coupling $\sim V$ commutes with the ring-exchange interaction, and a competition between both terms in the Hamiltonian (\ref{2Dmodel}) appears only in systems consisting of two or more plaquettes.  

\subsection{Spectroscopy}

While we are most interested in the dynamics induced by the Hamiltonian (\ref{2Dmodel}) in the gauge invariant subspace, we first describe an approach for probing signatures of the ring-exchange interaction \eqref{REhamiltonian} by performing spectroscopic measurements on the full circuit. To do so we assume that the four qubits can be individually coupled to a cavity resonator, which can be used to apply weak driving fields, as well as to detect microwave photons emitted from the qubits into the resonator. The resulting dynamics can be modeled by the master equation 
\beq \label{QuantumMasterEquation}
\dot{\rho} = - i [{H}+ H_{\rm drive}(t),{\rho}] + \frac{\Gamma}{2} \sum_{\ell} (2 {S}_{\ell}^{-} {\rho}{S}_{\ell}^{+} - \{ {S}_{\ell}^{+}{S}_{\ell}^{-}, {\rho} \} ),
\eeq
where the sum runs over all the spins (on the lattice links), $H$ is the Hamiltonian of Eq.~(\ref{MotherH}), and $H_{\rm drive}(t) = \sum_{\ell=1}^4 \Omega_\ell^d ( S_\ell^{+}e^{-i\omega_d t} + S_\ell^{-}e^{i\omega_d t})$ accounts for driving fields with frequency $\omega_d$ and site-dependent driving strength $\Omega_\ell^d$. In Eq.(\ref{QuantumMasterEquation}), $\Gamma$ is the qubit decay rate (assumed to be homogeneous), which limits the qubit performance and the accuracy of realizing gauge invariance. Under stationary driving conditions, the total number of photons emitted from a single qubit is proportional to the steady-state excited state population $\langle \sigma_{ee}(\ell)\rangle$, where $\sigma_{ee}(\ell) \equiv \frac{1}{2}+S_\ell^z$. By looking at correlated photon detection events, one also has access to functions of the form $\langle \sigma_{ee}(\ell)  \sigma_{ee}(\ell') \rangle$.

\begin{figure}
\centering
\includegraphics[width=\linewidth]{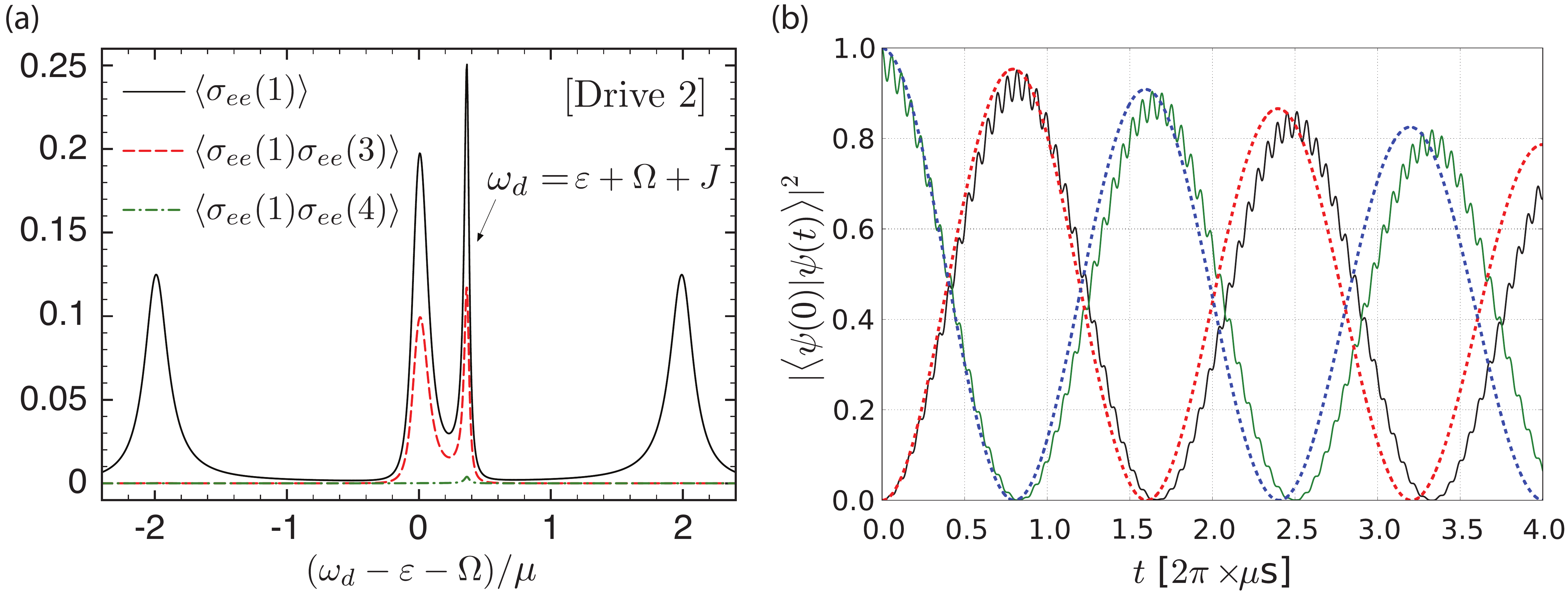}
\caption{
(Color online). (a) Excitation spectroscopy of a single plaquette [four coupled transmons -- see Fig.~\ref{plaquetteRing}(a)], as given by the microscopic Hamiltonian \eqref{MotherH}. Here the qubit $2$ is driven and the average population and correlations are computed in the steady state, considering qubit relaxation as captured by Eq.~(\ref{QuantumMasterEquation}). The population of qubit $1$ (solid line) yields a four-peak structure reminiscent of the one- and two-excitation subspaces shown in Fig.~\ref{plaquetteRing}(b). The peak at $\omega_d = \varepsilon+\Omega+J$ is a signature of the ring-exchange dynamics, as can be seen in cross-correlation measurements between different links (dashed lines). Here the parameter values are $\Gamma/(2\pi)=30$ kHz, $\Omega/(2\pi)=100$ MHz, $\mu/(2\pi)=7$ MHz, $\Omega_2^d/(2\pi) = 100$ kHz, $\Omega_{\ell\neq 2}^d = 0$. (b) Time-evolution of the gauge invariant states on a single plaquette given by the ring-exchange interaction \eqref{REhamiltonian}. Initially, the state $\ket{\downarrow \uparrow \downarrow \uparrow}$ (one excitation on the links 2 and 4) is prepared (lines starting at $|\langle\psi(0)|\psi(t)\rangle|^2=1$ for $t=0$). It coherently oscillates with the state $\ket{\uparrow \downarrow \uparrow \downarrow}$ (one excitation on the links 1 and 3). The microscopic model of Eq.~(\ref{MotherH}) (solid lines) is compared with the effective Hamiltonian (\ref{HGringEff}) (dotted lines). Including the effect of cavity decay, $\Gamma/(2\pi)=30$ kHz for all resonators, the population decays to $\sim 90\%$ after one oscillation. The values of the parameters are as above. Notice that these parameter values are not optimized --- in order to illustrate the feasibility under suboptimal conditions --- [c.f.\ Figs.~\ref{parameters} and \ref{twoPlaquettesDissipation}(a) for optimal parameter values]. Also note that the value of $\varepsilon$ is irrelevant for the effective dynamics.}
\label{population_drivings}
\end{figure}

Fig.~\ref{population_drivings}(a) shows the typical spectra in the case where qubit $2$ is driven in a single plaquette [c.f. Fig~\ref{plaquetteRing}(a)]. Measuring the excitation probability of the neighboring qubit $1$, $\langle \sigma_{ee}(1)\rangle$, one observes four distinct peaks, which can be identified with transitions between different energy eigenstates depicted in Fig.~\ref{plaquetteRing}(b). The two peaks at $\omega_d=\varepsilon+\Omega\pm 2\mu$ correspond to transitions from the ground state $|\downarrow\downarrow\downarrow\downarrow\rangle$ to eigenstates in the one-excitation manifold. Within this subspace, a single spin excitation can hop from site to site, thus forming delocalized eigenstates. The peak in the middle exhibits an additional splitting, which cannot be explained by the single excitation dynamics. It arises from a two-photon transition to the state $|\downarrow\uparrow\downarrow\uparrow\rangle$, which is then coupled to the state $|\uparrow\downarrow\uparrow\downarrow\rangle$ via the effective ring-exchange coupling, and leads to the characteristic splitting $\sim 2J$ of the transition.

Additional evidence for a correlated two-spin hopping interaction can be obtained from correlation measurements of the form $\langle \sigma_{ee}(\ell) \sigma_{ee}(\ell') \rangle$, which directly probe the two-excitation subspace. For example, as shown in Fig.~\ref{population_drivings}(a), the value of $\langle \sigma_{ee}(1) \sigma_{ee}(3) \rangle$ is no longer sensitive to the single excitation resonances, but still exhibits the ring-exchange splitting at $\omega_d\approx \varepsilon+\Omega$ and $\omega_d\approx \varepsilon+\Omega+J$. In contrast, the correlations between neighboring spins, e.g.\ $\langle \sigma_{ee}(1) \sigma_{ee}(4) \rangle$, vanish almost completely, since states of the type $|\uparrow\downarrow\downarrow \uparrow\rangle$ are not coupled via the ring-exchange Hamiltonian. Therefore, in combination, such measurements can be used to confirm that the relevant dynamics within the two-excitation subspace are indeed accurately described by the Hamiltonian (\ref{MotherH}), and thus --- effectively --- by (\ref{HGringEff}).

\subsection{Dynamics} 

In the remainder of the paper we are primarily interested in the dynamics induced by the effective Hamiltonian (\ref{HGringEff}), within the gauge invariant sector defined by $G_m|\psi\rangle=0$. For a single plaquette, this means that starting from the actual ground state of the circuit, $|\downarrow \downarrow \downarrow \downarrow \rangle$, at time $t=0$ we apply a fast microwave pulse to a selected set of qubits, which excite the system into one of the gauge invariant states, e.g.\ $|\uparrow \downarrow \uparrow\downarrow \rangle$. The subsequent dynamics is then given by the effective Hamiltonian, up to the point where one of the qubits decays. In Fig.~\ref{population_drivings}(b) we show the evolution given by both the microscopic Hamiltonian (\ref{MotherH}) and the effective model (\ref{HGringEff}) --- on a single plaquette --- including the effect of qubit decay. Preparing initially the state $\ket{\downarrow \uparrow \downarrow \uparrow}$ (one excitation on the links 2 and 4), this coherently oscillates with $\ket{\uparrow \downarrow \uparrow \downarrow}$ (one excitation on the links 1 and 3). Even for small qubit-qubit couplings, $\mu/(2\pi) \sim 7$ MHz, considered here, the microscopic model and the effective model agree qualitatively well, and start to be appreciably shifted only after a few oscillations. Assuming a qubit decay $\Gamma/(2\pi) = 30$ kHz \cite{Schoelkopf11, Steffen12}, the prepared-state population is $\sim 0.9$ after one oscillation, which shows the possibility of simulating the dynamics of the ring-exchange interaction with current superconducting circuits.

\section{Probing string dynamics} \label{2plaquettesSec}

\begin{figure}[t]
\begin{center}
\includegraphics[width=0.6\linewidth]{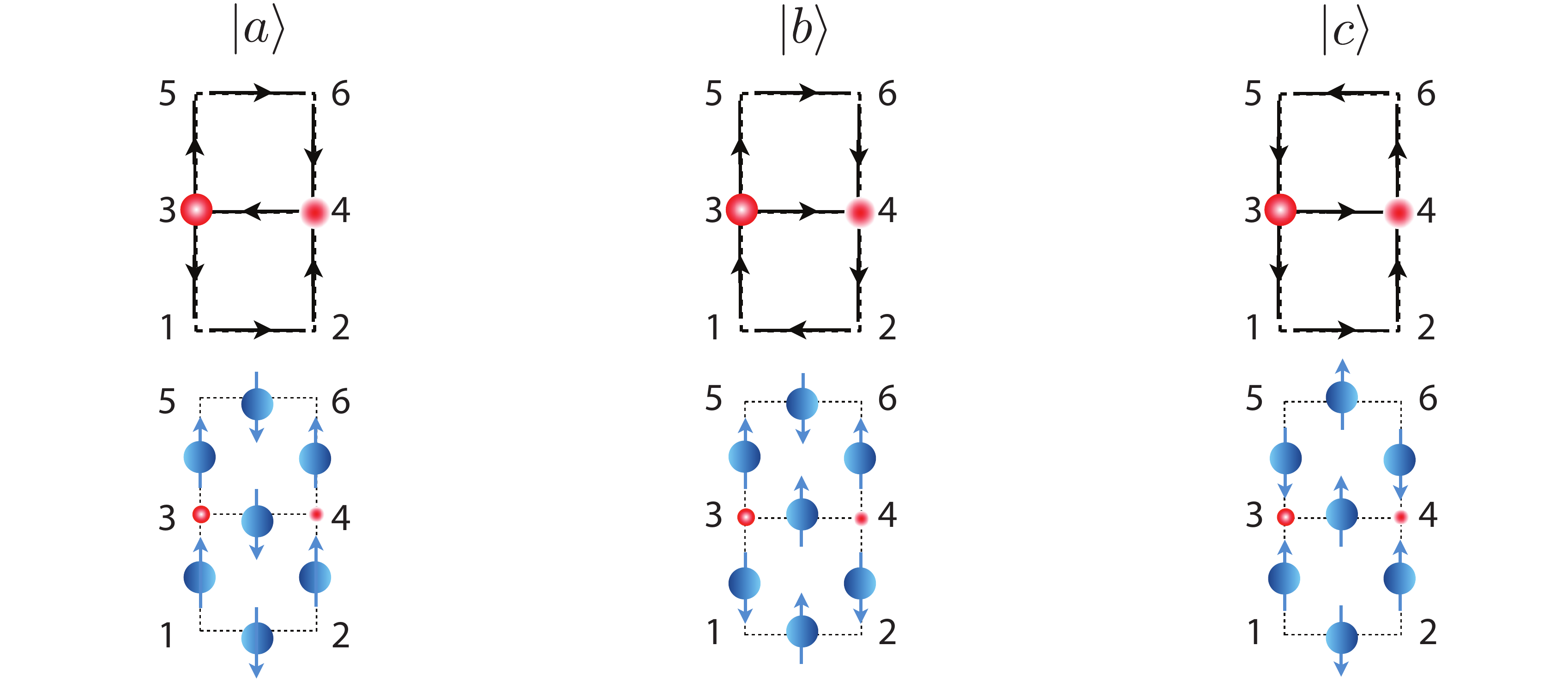}\\
\includegraphics[width=0.08\linewidth]{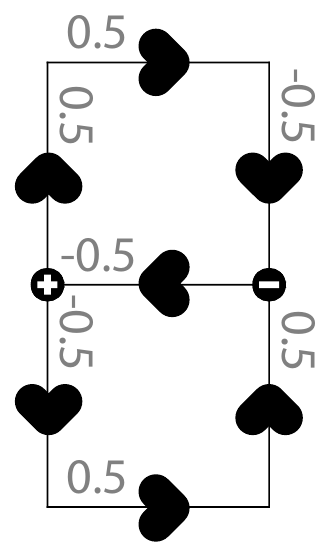} \qquad\qquad\qquad
\includegraphics[width=0.08\linewidth]{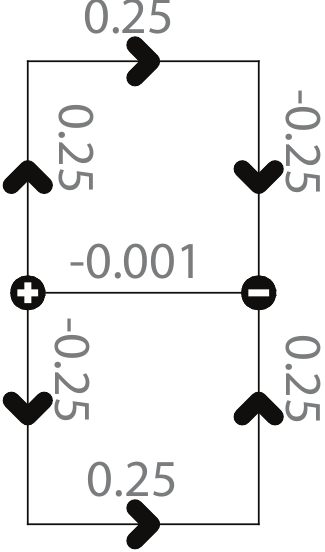}
\caption{(Color online). Upper panel: Flux configurations that obey the Gauss law on a lattice of two plaquettes, and corresponding spin configurations (below). Lower panel: Flux distribution in the ground state for $J/V = 0$ (left) and $J/V = -100$ (right). For $J/V = 0$ we recover the state $|a\rangle$, while for $|J/V| \gg 1$ the ground state is a superposition of the three gauge invariant states, with the electric flux propagating completely along the edges of the lattice.}
\label{Fig2}
\end{center}
\end{figure}

Next we discuss more complex dynamics given by the gauge invariant model (\ref{2Dmodel}). To this end we now consider the case $V \neq 0$, and study phenomena associated with the competing phases as the ratio $J/V$ is varied. First, we will show how the quantum phase transition of the model (\ref{2Dmodel}), present in the infinite-size limit, manifests itself as a crossover displayed by the magnetization of a single spin in a system of two plaquettes. Second, we study the physics of the electric flux strings connecting a charge and an anticharge in the lattice. 

In the upper panel of Fig.~\ref{Fig2} we show the possible configurations compatible with the Gauss law for a lattice of two plaquettes. Notice, that for a spin $\frac{1}{2}$ representation of the gauge fields, the Gauss law is irremediably broken at the vertices connected to three links. Therefore, in Fig.~\ref{Fig2} a charge-anticharge pair with $Q_m = \pm \frac{1}{2}$ has been created at the vertices $3$ and $4$. Furthermore, each of the states $|a\rangle$, $|b\rangle$, $|c\rangle$, is degenerate with the state corresponding to simultaneously inverting all the spins, a degeneracy that can be broken by applying a small magnetic field.
These states can be initially prepared by locally applying simultaneous $\pi$ pulses on the appropriate qubits. 
Starting e.g.~in $|a\rangle$, which corresponds to the ground state of the Hamiltonian $\eqref{2Dmodel}$ for $J=0$, $V>0$, we can adiabatically switch on the ring-exchange interaction to reach the ground state of the system for a particular ratio $J/V$. 
In the lower panel of Fig.~\ref{Fig2} we show a simulation of the ground-state flux distribution for $J/V =0$, $V>0$ (left) and for $J/V =-100$, $V>0$ (right). In the former case, the ground state is, as mentioned above, the antiferromagnetic state $|a\rangle$. However, when the ratio $|J/V|$ is increased, the ring-exchange term dominates the dynamics and the electric flux propagates from charge to anticharge along the edges of the lattice. In this case, the ground state is no longer a product state, but a quantum superposition of the states $|a\rangle$, $|b\rangle$, and $|c\rangle$.

\subsection{Finite-size crossover} \label{crossoverSec}

\begin{figure}[t]
\begin{center}
\includegraphics[width=\linewidth]{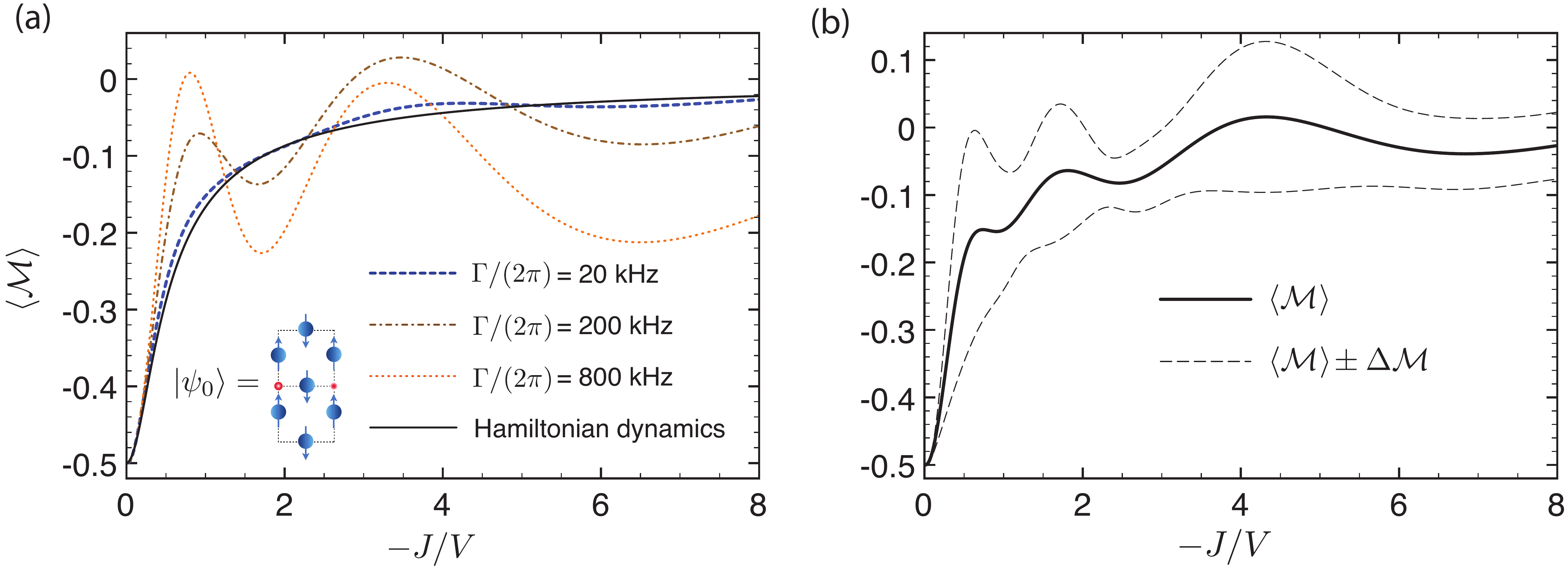}
\caption{(Color online). (a) The infinite-size quantum phase transition of the model (\ref{2Dmodel}) manifests itself as a crossover in a minimal lattice of two plaquettes. Here we have prepared the initial product state shown in the inset, and swept the parameters as $J = 30$ MHz$\times\sin^2(vt)$, $V = 30$ MHz$\times\cos^2(vt)$, with a constant speed $v/(2\pi) = 2\pi \times$ 2 MHz$/\mu$s. When the effect of qubit decay is considered, the spin on the common link [corresponding to the ``order parameter'' $\langle {\cal M} \rangle$ in this minimal case] decays at a rate $\Gamma$, thereby reducing the value of $\langle{\cal M}\rangle$ for large $|J/V|$. (b) Effect of disorder in a minimal lattice of two plaquettes. When the qubit frequencies take random values between $\pm \Delta \varepsilon/2$, the transition becomes less visible. Here we have taken $\Delta \varepsilon = 15$ MHz, and plotted the average $\langle {\cal M} \rangle$ (solid line) and standard deviation (dashed lines) over $10000$ realizations. The figure shows that, with uncertainties in the qubit frequencies of this magnitude, the crossover can still be observed. Here we have prepared the initial product state shown in the inset of Fig.~\ref{twoPlaquettesDissipation}(a), and swept the parameters as $J = 30$ MHz$\times\sin^2(vt)$, $V = 30$ MHz$\times\cos^2(vt)$, with a constant speed $v/(2\pi) = 2\pi \times$ 2 MHz$/\mu$s.}
\label{twoPlaquettesDissipation}
\end{center}
\end{figure}

In Fig.~\ref{twoPlaquettesDissipation}(a) we show how the infinite-size quantum phase transition of the model (\ref{2Dmodel}) manifests itself on a lattice of two plaquettes, captured by the average magnetization $\langle {\cal M} \rangle \equiv \langle S_{Q\bar{Q}}^z \rangle$ of the central spin between both plaquettes Here we start with the product state $|\psi_0\rangle = |a\rangle$ [c.f. Fig.~\ref{Fig2} and inset of Fig.~\ref{twoPlaquettesDissipation}(a)], which can be experimentally prepared by first cooling the system to the ground state \cite{GSqubitdecay} and then applying a simultaneous $\pi$ pulse on the appropriate links. We notice that this state is the ground state of the Hamiltonian (\ref{2Dmodel}) for $J=0$, $V>0$, and that the large energy scale $\sim \Omega$ ensures that the Gauss law is satisfied. In Fig.~\ref{twoPlaquettesDissipation}(a) we calculate $\langle {\cal M} \rangle$ when the parameters are varied with time as $J = J_0 \sin^2(vt)$, $V = V_0 \cos^2(vt)$, which, given a constant speed $v$, and amplitudes $J_0$, $V_0$, approximately follows the functional form shown in Fig.~\ref{parameters}(c). Neglecting qubit decay, $\langle {\cal M} \rangle$ increases from $-0.5$ to $0$. At finite relaxation rates, $\langle {\cal M} \rangle$ reaches a maximum at a finite value of $J/V$ and then decreases due to qubit decay. For standard relaxation rates [$\Gamma /(2\pi) \sim 20$ kHz] \cite{Schoelkopf11, Steffen12}, and superconducting-circuit parameters, the behavior of $\langle {\cal M} \rangle$ in the presence of qubit decay approximates well the one shown by the Hamiltonian dynamics, thereby allowing us to characterize the transition.

\subsubsection*{Disorder}

An important concern in the implementation of the model (\ref{2Dmodel}) is to what degree the crossover is masked by disorder (inhomogeneities among qubit frequencies). This effect is illustrated in Fig.~\ref{twoPlaquettesDissipation}(b), where we show the average of the magnetization, $\langle {\cal M} \rangle$, over $10000$ realizations (sufficient for convergence), with qubit frequencies taking random values between $\pm \Delta\varepsilon/2$. We notice that post-selecting qubits with similar frequencies after fabrication, or incorporating tuneable qubits, may allow uncertainties in qubit frequencies $\lesssim 15$ MHz (considered in Fig.~\ref{twoPlaquettesDissipation}(b)). For larger values of $\Delta \varepsilon$, the smoothening of the crossover shown in Fig.~\ref{twoPlaquettesDissipation}(b) becomes more pronounced, but up to $\Delta \varepsilon /(2\pi) \approx 50$ MHz, the crossover can still be well identified even in this small system. Although scaling to larger lattices leads to a higher probability of error (a common problem in quantum simulators) due to photon loss, a global order parameter such as the total magnetization is robust with respect to individual decay processes. Furthermore, a post-selection of measurements \cite{Fukuhara}, together with optimized pulses can be employed to increase the fidelity of the transition.

\subsection{String dynamics}

\begin{figure}
\begin{center}
\includegraphics[width=0.6\linewidth]{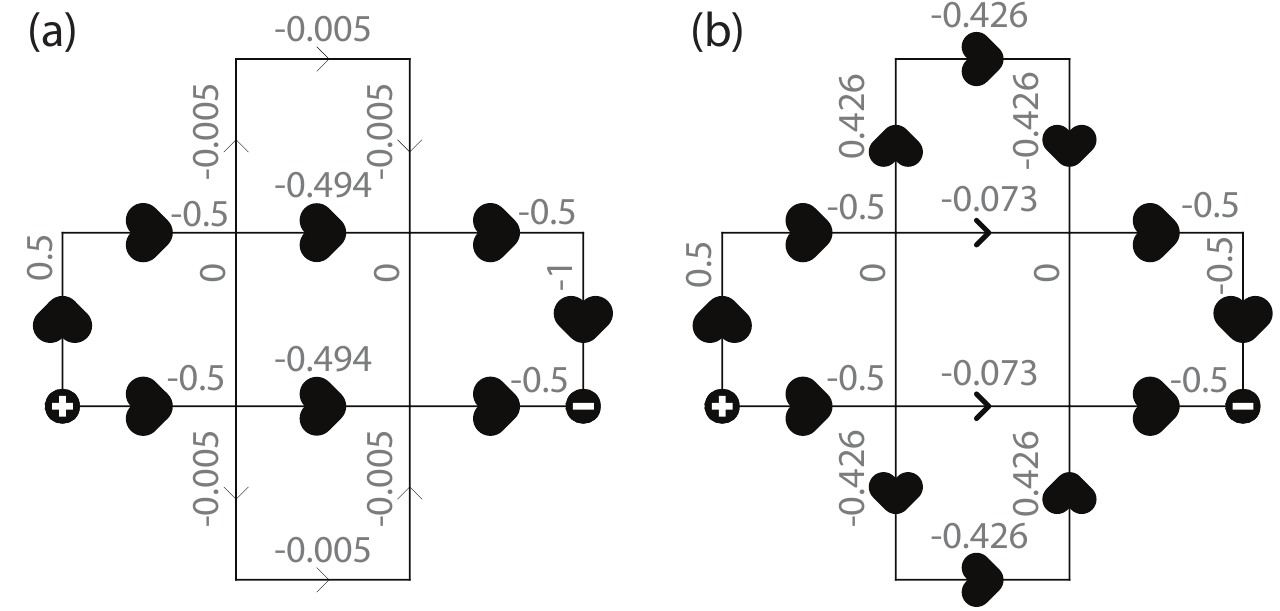}
\caption{Ground-state flux distribution in a lattice of five plaquettes. For $J/V = -1$ (a) the electric flux propagates from charge to anticharge through the center of the lattice, while for $J/V = 0.1$ (b) it propagates along the edges.}
\label{Fig5}
\end{center}
\end{figure}

As we have mentioned above, experimentally observing the dynamics of strings would give access to open questions about confinement in lattice gauge theories. In particular, performing time-resolved measurements would show the fluctuations of an initially-prepared string, and the formation of strands, a problem that, even for relatively small lattices, is challenging to simulate classically. In Fig.~\ref{Fig5} we show two particular examples of the ground-state distribution of flux, for a lattice of five plaquettes. Here we have created a charge-anticharge pair at the edges (achieved by a violation of the Gauss law by initially exciting/de-exciting the corresponding qubits). For $J/V = -1$ [Fig.~\ref{Fig5}(a)] the electric flux propagates from charge to anticharge mainly through the center of the lattice, while for $J/V = 0.1$ [Fig.~\ref{Fig5}(b)] it propagates along the edges of the system. This effect corresponds to a flux fractionalization into different strands, as it was observed in \cite{BanerjeeU1, BanerjeeArxiv14}. Experimentally, it would be interesting to investigate the time-dependence of this process, as well as the behavior as the ratio $J/V$ is varied.

\subsubsection*{Effect of dissipation on string dynamics}

\begin{figure}[t]
\begin{center}
\includegraphics[width=0.5\linewidth]{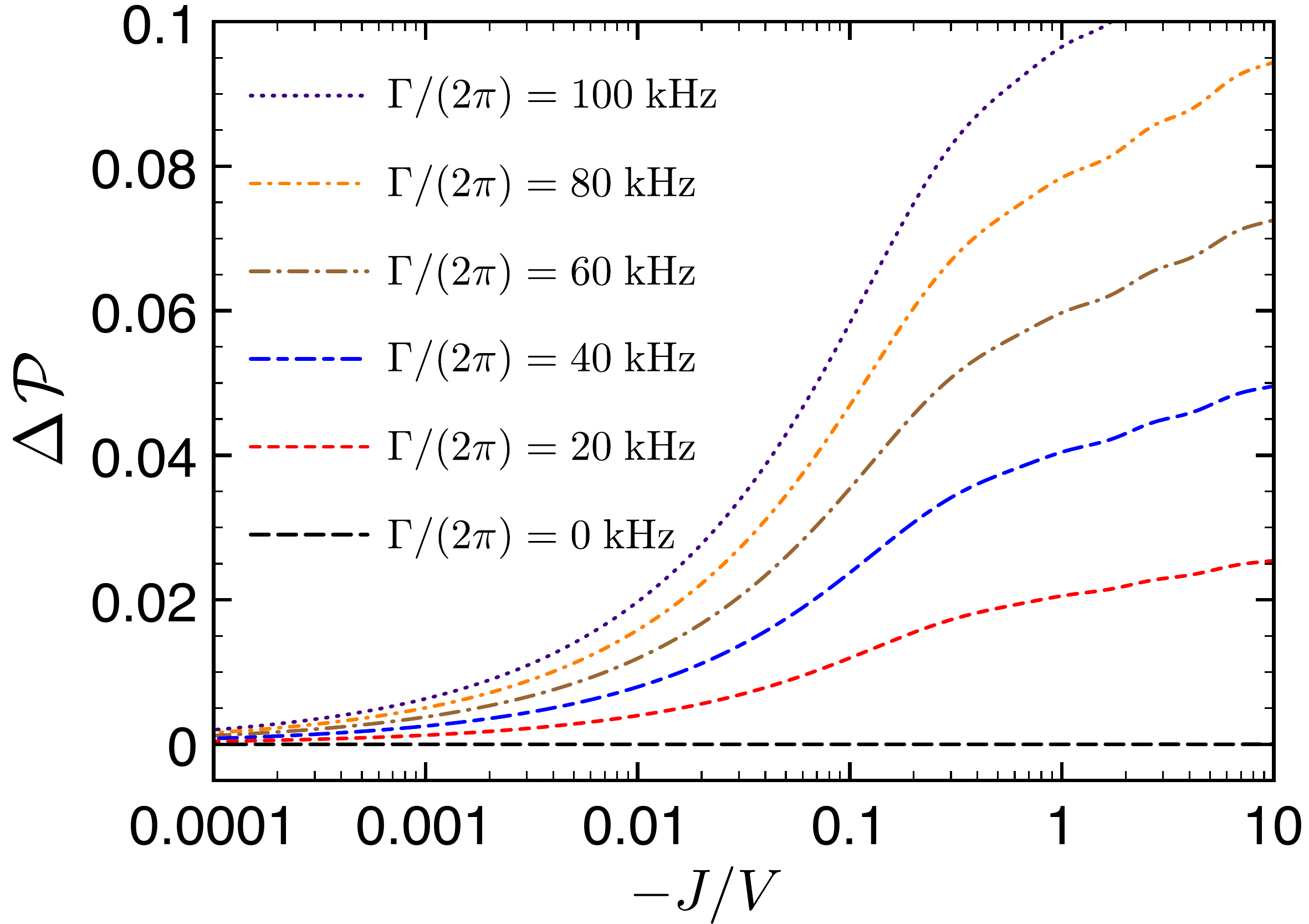}
\caption{(Color online). Effect of dissipation in a minimal lattice of two plaquettes [c.f. Fig.~\ref{Fig2}]. Here we have initially prepared the state $|a\rangle$ of Fig.~\ref{Fig2}, and swept the parameters according to $J = 30$ MHz$\times\sin^2(vt)$, $V = 30$ MHz$\times\cos^2(vt)$, with a constant speed $v/(2\pi) = 2\pi \times$ 2 MHz$/\mu$s. On the vertical axis we show $\Delta {\cal P} \equiv {\cal P}_{\Gamma=0} - {\cal P}_{\Gamma\neq 0}$ [see. Eq.~\eqref{probFidelity}] for different values of the qubit relaxation rate $\Gamma$. Although the probability of error to obtain the desired ground state for $J\neq 0$ increases with time due to excitation decay, it remains of the order of $2\%$ for realistic relaxation rates.}
\label{fidelity_dissipation}
\end{center}
\end{figure}

In order to measure the ground-state flux distribution shown in Fig.~\ref{Fig5}, an experimental protocol may consist on initially preparing a product state, ground state of the Hamiltonian \eqref{2Dmodel} for $J=0$, which corresponds to an antiferromagnet (ferromagnet) for $V>0$, ($V<0$). In the minimal lattice of Fig.~\ref{Fig2}, this initial gauge invariant configuration is precisely the state $|a\rangle$, where a string propagates from charge to anticharge along the edges of the lattice. In the lattice of Fig.~\ref{Fig5}, an equivalent string configuration --- compatible with the Gauss law --- can be initially prepared as a product state. By appropriately choosing the signs of the two-body Ising interactions in the lattice, this can be chosen equivalent to the ground state of the Hamiltonian $\eqref{2Dmodel}$. The highly-entangled ground state for $J\neq 0$ can then be reached by adiabatic evolution, with e.g. a sweep of the form shown in section \ref{crossoverSec}. During this protocol, it would be interesting to monitor the string dynamics as the ratio $J/V$ is varied. Notably, a common problem in quantum simulation is that the probability of reaching the appropriate ground state depends both on the system size and the qubit decoherence rates. This effect can be quantified by
\beq \label{probFidelity}
{\cal P} \equiv |\langle \psi_{\rm GS} (t) | \rho(t) | \psi_{\rm GS} (t) \rangle |,
\eeq
where $\rho(t)$ and $|\psi_{\rm GS}(t)\rangle$ are the system density operator and the ground-state wavefunction at time $t$, respectively. We can then define $\Delta {\cal P} \equiv {\cal P}_{\Gamma=0} - {\cal P}_{\Gamma\neq 0}$, which gives us the probability of error due to qubit decay.
Fig.~\ref{fidelity_dissipation} shows $\Delta {\cal P}$ for the system of Fig.~\ref{Fig2}, starting in the state $|a\rangle$, and during the adiabatic passage $J = J_0 \sin^2(vt)$, $V = V_0 \cos^2(vt)$, for a constant speed $v$, and amplitudes $J_0$, $V_0$. As the relaxation rate $\Gamma$ is increased, so does the probability of error during the transition. However, for state-of-the-art values, $\Gamma/(2\pi)\sim 20$ kHz, the probability of error to obtain the desired ground state at finite $J$ remains of the order of $2\%$.

\section{Conclusions and outlook} \label{ConclusionsSec}

In this work we have proposed an analog quantum simulator --- based on small-scale superconducting circuit lattices --- to engineer gauge invariant interactions. Specifically, we have shown how to construct ring-exchange couplings and four-body spin interactions in two spatial dimensions. Models involving such interactions are particularly relevant in the context of $U(1)$ quantum link models, quantum dimer physics, and spin ice. The characteristics of electric flux strings can be studied as well. This gives access to confinement properties, and to real-time dynamics in gauge invariant models. More generally, simulating gauge invariant interactions constitutes a toolbox to study open problems in quantum field theories.
As we have shown, for state-of-the-art superconducting circuits, and under realistic dissipative conditions, competing phases and the dynamics of confining strings can be investigated in small circuit lattices. The experimental realization of the gauge invariant models presented here, may thus address open questions in condensed matter and high-energy physics, and represents a first step towards the investigation of more complex interactions, such as the quantum simulation of non-Abelian gauge theories.

\section*{Acknowledgements}

We gratefully acknowledge discussions with D.\ Banerjee, M.\ Dalmonte, G.\ Kirchmair, and R.\ J.\ Schoelkopf. The research leading to these results has received funding from the Schweizerischer Na\-tio\-nal\-fonds and from the European Research Council under the European Union's Seventh Framework Programme (FP7/2007-2013)/ ERC grant agreement 339220. This work is also supported through the Austrian Academy of Sciences, the EU project SIQS, the ERC Synergy Grant UQUAM, the Austrian Science Fund (FWF) through SFB FOQUS, the START grant Y519-N16, the NSF-funded Physics Frontier Center at the JQI, and by the ARO MURI award W911NF0910406.

\appendix

\section{Eigenstates for one plaquette} \label{eigenstates_app}

The energies and eigenstates of the microscopic Hamiltonian \eqref{MotherH} for one plaquette are
\begin{equation}
\begin{split}
&E_0 = -2\varepsilon - \Omega, \quad |\psi_0 \rangle = |\downarrow\downarrow\downarrow\downarrow \rangle. \\
&E_1 = -\varepsilon - 2\mu, \quad |\psi_1 \rangle = ( |\downarrow\downarrow\downarrow\uparrow\rangle - | \downarrow\downarrow\uparrow\downarrow \rangle + | \downarrow\uparrow\downarrow\downarrow \rangle - |\uparrow\downarrow\downarrow\downarrow\rangle )/2. \\
&E_2 = -\varepsilon, \quad |\psi_2\rangle \approx ( |\uparrow\downarrow\downarrow\downarrow \rangle - | \downarrow\downarrow\uparrow\downarrow \rangle )/\sqrt{2}. \\
&E_3 = - \varepsilon, \quad |\psi_3 \rangle = (|\downarrow\uparrow\downarrow\downarrow \rangle - |\downarrow\downarrow\downarrow\uparrow \rangle)/\sqrt{2}. \\
&E_4 = -\varepsilon+2\mu, \quad |\psi_4\rangle = ( |\downarrow\downarrow\downarrow\uparrow\rangle + | \downarrow\downarrow\uparrow\downarrow \rangle + | \downarrow\uparrow\downarrow\downarrow \rangle + |\uparrow\downarrow\downarrow\downarrow\rangle )/2. \\
&E_5 = \frac{1}{2} (\Omega-\sqrt{\Omega^2+32\mu^2}) \approx -2J, \quad |\psi_5\rangle \approx - \frac{1}{2\sqrt{1+8(\mu/\Omega)^2}} (|\uparrow\uparrow\downarrow\downarrow\rangle + | \uparrow\downarrow\downarrow\uparrow \rangle + | \downarrow\downarrow\uparrow\uparrow \rangle) \nonumber\\ &\; \quad\quad\quad\quad\quad\quad\quad\quad\quad\quad\quad\quad\quad\quad\quad\quad \;\; + \frac{(\mu/\Omega)^2}{2\sqrt{1+8(\mu/\Omega)^2}} (|\uparrow\downarrow\uparrow\downarrow\rangle + |\downarrow\uparrow\downarrow\uparrow\rangle). \\
&E_6 = 0, \quad |\psi_6\rangle \approx (|\uparrow\uparrow\downarrow\downarrow\rangle - |\uparrow\downarrow\downarrow\uparrow \rangle)/\sqrt{2}. \\
&E_7 = 0, \quad |\psi_7\rangle \approx (|\uparrow\uparrow\downarrow\downarrow\rangle - |\downarrow\uparrow\uparrow\downarrow \rangle)/\sqrt{2}. \\
&E_8 = 0, \quad |\psi_8\rangle \approx (|\uparrow\uparrow\downarrow\downarrow\rangle - |\downarrow\downarrow\uparrow\uparrow \rangle)/\sqrt{2}. \\
&E_9 = \Omega, \quad |\psi_9\rangle \approx (|\uparrow\downarrow\uparrow\downarrow \rangle - |\downarrow\uparrow\downarrow\uparrow \rangle)/\sqrt{2}. \\
&E_{10} = \frac{1}{2} (\Omega+\sqrt{\Omega^2+32\mu^2}) \approx \Omega + 2 J, \quad |\psi_{10}\rangle \approx \Big[ 1/\sqrt{2} - 2\sqrt{2} (\mu/\Omega)^2 \Big] (|\uparrow\downarrow\uparrow\downarrow\rangle + |\downarrow\uparrow\downarrow\uparrow \rangle) \nonumber\\ &\; \quad\quad\quad\quad\quad\quad\quad\quad\quad\quad\quad\quad\quad\quad\quad\quad\quad \;\;\; + \sqrt{2}(\mu/\Omega) (| \uparrow\uparrow\downarrow\downarrow \rangle + | \uparrow\downarrow\downarrow\uparrow \rangle + | \downarrow\uparrow\uparrow\downarrow \rangle + | \downarrow\downarrow\uparrow\uparrow \rangle). \\
&E_{11} = \varepsilon - 2\mu, \quad |\psi_{11} \rangle = (|\uparrow\uparrow\uparrow\downarrow \rangle - |\uparrow\uparrow\downarrow\uparrow \rangle + |\uparrow\downarrow\uparrow\uparrow \rangle - |\downarrow\uparrow\uparrow\uparrow \rangle)/2. \\
&E_{12} = \varepsilon, \quad |\psi_{12} \rangle \approx (|\uparrow\downarrow\uparrow\uparrow \rangle - | \uparrow\uparrow\uparrow\downarrow \rangle)/\sqrt{2}. \\
&E_{13} = \varepsilon, \quad |\psi_{13} \rangle = (|\downarrow\uparrow\uparrow\uparrow \rangle - |\uparrow\uparrow\downarrow\uparrow \rangle)/\sqrt{2}. \\
&E_{14} = \varepsilon + 2\mu, \quad |\psi_{14}\rangle = - (|\uparrow\uparrow\uparrow\downarrow\rangle + |\uparrow\uparrow\downarrow\uparrow \rangle + |\uparrow\downarrow\uparrow\uparrow \rangle + |\downarrow\uparrow\uparrow\uparrow \rangle)/\sqrt{2}. \\
&E_{15} = 2\varepsilon -\Omega, \quad |\psi_{15}\rangle = |\uparrow\uparrow\uparrow\uparrow \rangle.
\end{split}
\end{equation}


\newpage
 
\begin{thebibliography}{99}

\bibitem{Nakamura}
Y. Nakamura, Yu. A. Pashkin, and J. S. Tsai, \href{http://www.nature.com/nature/journal/v398/n6730/full/398786a0.html} {Nature {\bf 398}, 786 (1999).}

\bibitem{vanderWal}
C. H. van der Wal, A. C. J. ter Haar, F. K. Wilhelm, R. N. Schouten, C. J. P. M. Harmans, T. P. Orlando, S. Lloyd, and J. E. Mooij, \href{http://www.sciencemag.org/content/290/5492/773.abstract} {Science {\bf 290}, 773 (2000).}

\bibitem{Wallraff13}
L. Steffen, Y. Salathe, M. Oppliger, P. Kurpiers, M. Baur, C. Lang, C. Eichler, G. Puebla-Hellmann, A. Fedorov, and A. Wallraff, \href{http://www.nature.com/nature/journal/v500/n7462/full/nature12422.html} {Nature {\bf 500}, 319 (2013).}

\bibitem{DiCarlo09}
L. DiCarlo, J. M. Chow, J. M. Gambetta, L. S. Bishop, B. R. Johnson, D. I. Schuster, J. Majer, A. Blais, L. Frunzio, S. M. Girvin, and R. J. Schoelkopf, \href{http://www.nature.com/nature/journal/v460/n7252/full/nature08121.html} {Nature {\bf 460}, 240 (2009).}

\bibitem{Martinis12}
E. Lucero, R. Barends, Y. Chen, J. Kelly, M. Mariantoni, A. Megrant, P. OÕMalley, D. Sank, A. Vainsencher, J. Wenner, T. White, Y. Yin,	 A. N. Cleland, and J. M. Martinis, \href{http://www.nature.com/nphys/journal/v8/n10/full/nphys2385.html} {Nature Phys. {\bf 8}, 719 (2012).}

\bibitem{Martinis11}
M. Mariantoni, H. Wang, T. Yamamoto, M. Neeley, R. C. Bialczak, Y. Chen, M. Lenander, E. Lucero, A. D. O'Connell, D. Sank, M. Weides, J. Wenner, Y. Yin, J. Zhao, A. N. Korotkov, A. N. Cleland, and J M. Martinis, \href{http://www.sciencemag.org/content/334/6052/61.abstract} {Science {\bf 334}, 61 (2011).}

\bibitem{Wallraff12}
A. Fedorov, L. Steffen, M. Baur,	 M. P. da Silva, and A. Wallraff, \href{http://www.nature.com/nature/journal/v481/n7380/full/nature10713.html} {Nature {\bf 481}, 170 (2012).}

\bibitem{Schoelkopf12}
M. D. Reed, L. DiCarlo, S. E. Nigg, L. Sun, L. Frunzio, S. M. Girvin, and R. J. Schoelkopf, \href{http://www.nature.com/nature/journal/v482/n7385/full/nature10786.html} {Nature {\bf 482}, 382 (2012).}

\bibitem{Schoelkopf11}
H. Paik, D. I. Schuster, L. S. Bishop, G. Kirchmair, G. Catelani, A. P. Sears, B. R. Johnson, M. J. Reagor, L. Frunzio, L. I. Glazman, S. M. Girvin, M. H. Devoret, and R. J. Schoelkopf, \href{http://prl.aps.org/abstract/PRL/v107/i24/e240501} {Phys. Rev. Lett. {\bf 107}, 240501 (2011).}

\bibitem{Steffen12}
C. Rigetti, J. M. Gambetta, S. Poletto, B. L. T. Plourde, J. M. Chow, A. D. C\'orcoles, J. A. Smolin, S. T. Merkel, J. R. Rozen, G. A. Keefe, M. B. Rothwell, M. B. Ketchen, and M. Steffen, \href{http://prb.aps.org/abstract/PRB/v86/i10/e100506} {Phys. Rev. B {\bf 86}, 100506(R) (2012).}

\bibitem{Kirchmair13}
G. Kirchmair, B. Vlastakis,	 Z. Leghtas, S. E. Nigg, H. Paik,	 E. Ginossar, M. Mirrahimi, L. Frunzio, S. M. Girvin, and R. J. Schoelkopf, \href{http://www.nature.com/nature/journal/v495/n7440/full/nature11902.html} {Nature {\bf 495}, 205 (2013).}

\bibitem{DevoretSchoelkopfReview}
M. H. Devoret and R. J. Schoelkopf, \href{http://www.sciencemag.org/content/339/6124/1169.full} {Science {\bf 339}, 1169 (2013).}

\bibitem{HouckTureciKoch}
A. A. Houck, H. E. T\"ureci, and J. Koch, \href{http://www.nature.com/nphys/journal/v8/n4/abs/nphys2251.html}{Nature Phys. {\bf 8}, 292 (2012).}

\bibitem{Houck12}
D. L. Underwood, W. E. Shanks, J. Koch, and A. A. Houck, \href{http://pra.aps.org/abstract/PRA/v86/i2/e023837} {Phys. Rev. A {\bf 86}, 023837 (2012).}

\bibitem{Koch10}
J. Koch, A. A. Houck, K. Le Hur, and S. M. Girvin, \href{http://pra.aps.org/abstract/PRA/v82/i4/e043811} {Phys. Rev. A {\bf 82}, 043811 (2010).}

\bibitem{Preskill}
S. P. Jordan, K. S. M. Lee, and J. Preskill, \href{http://www.sciencemag.org/content/336/6085/1130}{Science {\bf 336}, 1130 (2012).}

\bibitem{WieseReview}
U.-J. Wiese, \href{http://onlinelibrary.wiley.com/doi/10.1002/andp.201300104/full} {Ann. Phys. {\bf 525}, 777 (2013).}

\bibitem{Wilson}
K. G. Wilson, \href{http://prd.aps.org/abstract/PRD/v10/i8/p2445_1} {Phys. Rev. D {\bf 10}, 2445 (1974).} 

\bibitem{Kogut-Susskind}
J. Kogut and L. Susskind, \href{http://prd.aps.org/abstract/PRD/v11/i2/p395_1} {Phys. Rev. D {\bf 11}, 395 (1975).}

\bibitem{Gattringer}
C. Gattringer and C. B. Lang, \href{http://books.google.at/books/about/Quantum_Chromodynamics_on_the_Lattice.html?id=l2hZKnlYDxoC&redir_esc=y} {{\it Quantum Chromodynamics on the Lattice} (Springer-Verlag, Berlin Heidelberg, 2010).}

\bibitem{KogutSpinsRMP}
J. B. Kogut, \href{http://rmp.aps.org/abstract/RMP/v51/i4/p659_1} {Rev. Mod. Phys. {\bf  51}, 659 (1979).}

\bibitem{Wen} 
X.-G. Wen, {\it Quantum Field Theory of Many-body Systems} (Oxford University Press, New York, 2004).

\bibitem{LeeNagaosaWen}
P. A. Lee, N. Nagaosa, and X.-G. Wen, \href{http://journals.aps.org/rmp/abstract/10.1103/RevModPhys.78.17}{Rev. Mod. Phys. {\bf 78}, 17 (2006).}

\bibitem{Lacroix}
C. Lacroix, P. Mendels, and F. Mila, \href{http://books.google.at/books/about/Introduction_to_Frustrated_Magnetism.html?id=utSV09ZuhOkC&redir_esc=y} {\it Introduction to Frustrated Magnetism} (Springer-Verlag, Berlin, 2011).

\bibitem{Balents}
L. Balents, \href{http://www.nature.com/nature/journal/v464/n7286/full/nature08917.html} {Nature {\bf 464}, 199 (2010).}

\bibitem{Buchler05}
H. P. B\"uchler, M. Hermele, S. D. Huber, M. P. A. Fisher, and P. Zoller, \href{http://journals.aps.org/prl/abstract/10.1103/PhysRevLett.95.040402} {Phys. Rev. Lett. {\bf 95}, 040402 (2005).}

\bibitem{Weimer10}
H. Weimer, M. M\"uller, I. Lesanovsky, P. Zoller, and H. P. B\"uchler, \href{http://www.nature.com/nphys/journal/v6/n5/abs/nphys1614.html} {Nature Phys. {\bf 6}, 382 (2010).}

\bibitem{Kapit11}
E. Kapit and E. Mueller, \href{http://pra.aps.org/abstract/PRA/v83/i3/e033625} {Phys. Rev. A {\bf 83}, 033625 (2011).}

\bibitem{Zohar11}
E. Zohar and B. Reznik, \href{http://prl.aps.org/abstract/PRL/v107/i27/e275301} {Phys. Rev. Lett {\bf 107}, 275301 (2011).}

\bibitem{Banerjee12}
D. Banerjee, M. Dalmonte, M. M\"uller, E. Rico, P. Stebler, U.-J. Wiese, and P. Zoller, \href{http://prl.aps.org/abstract/PRL/v109/i17/e175302} {Phys. Rev. Lett. {\bf 109}, 175302 (2012).}

\bibitem{Zohar12}
E. Zohar, J. I. Cirac, and B. Reznik, \href{http://prl.aps.org/abstract/PRL/v109/i12/e125302} {Phys. Rev. Lett. {\bf 109}, 125302 (2012).}

\bibitem{Zohar13}
E. Zohar, J. I. Cirac, and B. Reznik, \href{http://prl.aps.org/abstract/PRL/v110/i5/e055302} {Phys. Rev. Lett. {\bf 110}, 055302 (2013).}

\bibitem{Lewenstein12}
L. Tagliacozzo, A. Celi, A. Zamora, and M. Lewenstein, \href{http://www.sciencedirect.com/science/article/pii/S0003491612001819} {Ann. Phys. {\bf 330}, 160 (2013).}

\bibitem{Banerjee13}
D. Banerjee, M. B\"ogli, M. Dalmonte, E. Rico, P. Stebler, U.-J. Wiese, and P. Zoller, \href{http://prl.aps.org/abstract/PRL/v110/i12/e125303} {Phys. Rev. Lett. {\bf 110}, 125303 (2013).}

\bibitem{Zohar13b}
E. Zohar, J. I. Cirac, and B. Reznik, \href{http://prl.aps.org/abstract/PRL/v110/i12/e125304} {Phys. Rev. Lett. {\bf 110}, 125304 (2013).}

\bibitem{Lewenstein13}
L. Tagliacozzo, A. Celi, P. Orland, M. W. Mitchell, and M. Lewenstein, \href{http://www.nature.com/ncomms/2013/131028/ncomms3615/full/ncomms3615.html} {Nature Comm. doi:10.1038/ ncomms3615.}

\bibitem{Zohar13c}
E. Zohar, J. I. Cirac, and B. Reznik, \href{http://pra.aps.org/abstract/PRA/v88/i2/e023617} {Phys. Rev. A {\bf 88}, 023617 (2013).}

\bibitem{Glaetzle}
A. W. Glaetzle, M. Dalmonte, R. Nath, I. Rousochatzakis, R. Moessner, and P. Zoller, \href{http://arxiv.org/abs/1404.5326}{arXiv:1404.5326.}

\bibitem{Ioffe}
L. B. Ioffe, M. V. Feigel'man, A. Ioselevich, D. Ivanov, M. Troyer, and G. Blatter, \href{http://www.nature.com/nature/journal/v415/n6871/full/415503a.html} {Nature {\bf 415}, 503 (2002).}

\bibitem{Hauke13}
P. Hauke, D. Marcos, M. Dalmonte, and P. Zoller, \href{http://journals.aps.org/prx/abstract/10.1103/PhysRevX.3.041018} {Phys. Rev. X {\bf 3}, 041018 (2013).}

\bibitem{Marcos13}
D. Marcos, P. Rabl, E. Rico, and P. Zoller, \href{http://prl.aps.org/abstract/PRL/v111/i11/e110504} {Phys. Rev. Lett. {\bf 111}, 110504 (2013).}

\bibitem{Horn81} 
D. Horn, \href{http://www.sciencedirect.com/science/article/pii/0370269381907632} {Phys. Lett. B {\bf 100}, 149 (1981).}

\bibitem{Orland90}
P. Orland and D. Rohrlich, \href{http://www.sciencedirect.com/science/article/pii/055032139090646U} {Nuc. Phys. B, {\bf 338}, 647 (1990).}

\bibitem{Wiese97}
S. Chandrasekharan and U.-J Wiese, \href{http://www.sciencedirect.com/science/article/pii/S0550321397800417} {Nuc. Phys. B, {\bf 492}, 455 (1997).}

\bibitem{Bro99}
R. Brower, S. Chandrasekharan, and U.-J. Wiese, \href{http://journals.aps.org/prd/abstract/10.1103/PhysRevD.60.094502}{Phys. Rev. D {\bf 60}, 094502 (1999).}

\bibitem{AndersonRVB}
P. W. Anderson, \href{http://www.sciencemag.org/content/235/4793/1196.abstract}{Science {\bf 235}, 1196 (1987).}

\bibitem{BanerjeeU1}
D. Banerjee, F.-J. Jiang, P. Widmer, and U.-J. Wiese, \href{http://iopscience.iop.org/1742-5468/2013/12/P12010}{J. Stat. Mech. (2013) P12010.}

\bibitem{BanerjeeProceedings}
D. Banerjee, P. Widmer, F.J. Jiang and U.-J. Wiese, \href{http://pos.sissa.it/cgi-bin/reader/conf.cgi?confid=187} {PoS (LATTICE 2013) 333.}

\bibitem{BanerjeeArxiv14}
D. Banerjee, M. B\"ogli, C. P. Hofmann, F.-J. Jiang, P. Widmer, U.-J. Wiese, \href{http://arxiv.org/abs/1406.2077}{arXiv:1406.2077.}

\bibitem{Trotzky}
S. Trotzky, Y-A. Chen, A. Flesch, I. P. McCulloch, U. Schollw\"ock, J. Eisert, and I. Bloch, \href{http://www.nature.com/nphys/journal/v8/n4/full/nphys2232.html}{Nature Phys. {\bf 8}, 325 (2012).}

\bibitem{RokhsarKivelson}
D. S. Rokhsar and S. A. Kivelson, \href{http://prl.aps.org/abstract/PRL/v61/i20/p2376_1} {Phys. Rev. Lett. {\bf 61}, 2376 (1998).}

\bibitem{KochTransmon}
J. Koch, T. M. Yu, J. Gambetta, A. A. Houck, D. I. Schuster, J. Majer, A. Blais, M. H. Devoret, S. M. Girvin, and R. J. Schoelkopf, \href{http://pra.aps.org/abstract/PRA/v76/i4/e042319} {Phys. Rev. A {\bf 76}, 042319 (2007).}

\bibitem{Schreier}
J. A. Schreier, A. A. Houck, J. Koch, D. I. Schuster, B. R. Johnson, J. M. Chow, J. M. Gambetta, J. Majer, L. Frunzio, M. H. Devoret, S. M. Girvin, and R. J. Schoelkopf, \href{http://journals.aps.org/prb/abstract/10.1103/PhysRevB.77.180502}{Phys. Rev. B {\bf 77}, 180502(R) (2008).}

\bibitem{Tian}
A.V. Sharypov, X. Deng, and L. Tian, \href{http://journals.aps.org/prb/abstract/10.1103/PhysRevB.86.014516}{Phys. Rev. B {\bf 86}, 014516 (2012).}

\bibitem{Hartmann13}
J. Jin, D. Rossini, R. Fazio, M. Leib, and M. J. Hartmann, \href{http://journals.aps.org/prl/abstract/10.1103/PhysRevLett.110.163605}{Phys. Rev. Lett. {\bf 110}, 163605 (2013).}

\bibitem{GSqubitdecay}
Notice that, when considering the effect of qubit decay, the ground state corresponds to all the spins pointing down.

\bibitem{Fukuhara}
T. Fukuhara, A. Kantian, M. Endres, M. Cheneau, P. Schau§, S. Hild, D. Bellem, U. Schollw\"ock, T. Giamarchi, C. Gross, I. Bloch, and S. Kuhr, \href{http://www.nature.com/nphys/journal/v9/n4/full/nphys2561.html}{Nature Phys. {\bf 9}, 235 (2013).}

\end{thebibliography}
\end{document}